\documentclass[amsmath,aip,jcp,reprint,superscriptaddress,twocolumn]{revtex4-1}

\usepackage{color}
\usepackage{amsfonts}
\usepackage{amssymb}
\usepackage{graphicx}

\begin{document}
\title{Perspective: Green's function methods for single molecule junctions}
\author{Guy Cohen}
\email{gcohen@tau.ac.il}
\affiliation{The Raymond and Beverley Sackler Center for Computational Molecular and Materials Science, Tel Aviv University, Tel Aviv 69978, Israel}
\affiliation{School of Chemistry, Tel Aviv University, Tel Aviv 69978, Israel}
\author{Michael Galperin}
\email{migalperin@ucsd.edu}
\affiliation{Department of Chemistry \& Biochemistry, University of California San Diego, La Jolla, CA 92093, USA}

\begin{abstract}
We present a brief pedagogical review of theoretical Green's function methods applicable to open quantum systems out of equilibrium in general, and single molecule junctions in particular.
We briefly describe experimental advances in molecular electronics, then discuss different theoretical approaches.
We then focus on Green's function methods.
Two characteristic energy scales governing the physics are many-body interactions within the junctions, and molecule--contact coupling.
We therefore discuss weak interactions and weak coupling, as two limits that can be conveniently treated within, respectively, the standard nonequilibrium Green's function (NEGF) method and its many-body flavors (pseudoparticle and Hubbard NEGF).
We argue that the intermediate regime, where the two energy scales are comparable, can in many cases be efficiently treated within the recently introduced superperturbation dual fermion approach.
Finally, we review approaches for going beyond these analytically accessible limits, as embodied by recent developments in numerically exact methods based on Green's functions.
\end{abstract}

\maketitle

\section{Introduction}\label{intro}
Since the first theoretical proposal to use single molecules as electronic devices \cite{aviram_molecular_1974}
and the first experimental realization of a single molecule junction,\cite{reed_conductance_1997,reed_computing_2000}
the field of molecular electronics has made tremendous progress.
This is evidenced by improved nanoscale fabrication techniques
and a significant increase in the variety of signals measurable in single molecule junctions. \cite{ratner_brief_2013}
As experimental techniques developed, the focus of research has shifted over the years from measurements of elastic coherent transport in junctions \cite{reed_conductance_1997,langlais_spatially_1999,smit_measurement_2002} 
to studies of quantum coherence effects  \cite{mayor_electric_2003,vazquez_probing_2012,arroyo_signatures_2013,borges_role_2017}
and to improving the stability and reproducibility of measurements. \cite{cohen_direct_2005,tao_electron_2006,venkataraman_dependence_2006,venkataraman_single-molecule_2006,cheng_situ_2011} 

Increasingly sensitive measurements allowed for detection of inelastic effects in the off-resonant tunneling regime,
leading to the appearance of inelastic electron tunneling spectroscopy (IETS). \cite{hahn_electronic_2000,wang_inelastic_2004,tal_molecular_2009,okabayashi_inelastic_2010,huan_spatial_2011,chiang_real-space_2014,yu_insights_2015,li_rotational_2015,schwarz_charge_2016,han_probing_2017,czap_detection_2019}
IETS is particularly useful as a diagnostic tool, since it provides vibrational fingerprints of molecules within the junction. \cite{wang_inelastic_2004}
It is also instrumental for imaging of molecular structure and chemical bonding,\cite{chiang_real-space_2014,li_rotational_2015}
as well as for probing intermolecular interactions and anharmonic overtones.\cite{han_probing_2017,czap_detection_2019}

Inelastic measurements in the resonant regime (RIETS) can also be performed, where the inelastic signal reveals itself as a peak in conductance that indicates positions of vibronic molecular levels.\cite{park_nanomechanical_2000,leroy_electrical_2004,franke_excitation_2010,gaudenzi_transport_2017}
The electronic population fluctuations within the molecule in this regime lead to stronger electron--vibration interactions, which then leads to destruction of coherence\cite{ballmann_experimental_2012} and to reorganization of polarization (polaron formation) in the junction.\cite{he_electrochemical_2006}
The former also controls the transition from tunneling-dominated transport to the hopping regime,\cite{lu_tunneling_2009,germs_unusual_2012,nichols_single_2016,taherinia_charge_2016,li_thermoelectric_2016,morteza_najarian_structure_2017,merces_long-range_2017,nguyen_highly_2018,miwa_hubbard_2019,ie_highly_2019,poggini_temperature-induced_2019} while the latter may result in, e.g., conformational changes in molecular structure.
This has been considered as a possible mechanism for highly nonlinear current--voltage characteristics such as negative differential resistance (NDR) and hysteresis. \cite{chen_large_1999,rawlett_electrical_2002,li_fabrication_2003,blum_molecularly_2005,he_electrochemical_2006,lortscher_reversible_2006,keane_three-terminal_2006,kiehl_charge_2006,wu_conductance_2008,osorio_conductance_2010,schwarz_field-induced_2016,fung_breaking_2019}  

A related body of research involves molecular heating/cooling driven by the electronic flux \cite{schulze_resonant_2008,lee_heat_2013,cui_peltier_2018} and thermoelectricity.\cite{ludoph_thermopower_1999,kim_thermal_2001,reddy_thermoelectricity_2007,widawsky_simultaneous_2012,meier_length-dependent_2014,kim_electrostatic_2014,kim_radiative_2015,cui_perspective:_2017}
Heating is directly relevant to questions regarding junction stability and the reproducibility of measurements.
Nanoscale thermoelectrics raises interesting potential technological possibilities, such as utilization of quantum effects \cite{schwab_measurement_2000,cui_quantized_2017} for constructing highly efficient nanoscale devices for energy transduction.

Historically, IETS was the first spectroscopic tool to be applied to single molecule junctions.
However, developments in laser techniques also allowed for performing
optical spectroscopy measurements in current-carrying molecular junctions.
This combination of optical spectroscopy and molecular electronics lead to the appearance of a new field of research coined molecular optoelectronics. \cite{galperin_molecular_2012}
In particular, surface enhanced Raman spectroscopy (SERS) \cite{ward_simultaneous_2008,shamai_spectroscopy_2011,natelson_nanogap_2013,li_voltage_2014} 
in junctions, besides providing (complementary to IETS) information on molecular vibrations, allows for estimating bias-induced heating of electronic and vibrational degrees of freedom. \cite{ioffe_detection_2008,ward_vibrational_2011}
Tip-enhanced Raman spectroscopy (TERS) \cite{jiang_observation_2012,sonntag_origin_2013} 
yields information on molecular structure. \cite{liu_revealing_2011,zhang_chemical_2013,chiang_molecular-resolution_2015,lee_tip-enhanced_2017,sun_role_2019}
Furthermore, optical emission in biased junctions was observed as bias-induced luminescence. \cite{qiu_vibrationally_2003,dong_vibrationally_2004,wu_intramolecular_2008,seldenthuis_electroluminescence_2010,braun_superluminescence_2015,fung_too_2017}
In this context, electroluminescence was employed to study energy transfer \cite{imada_real-space_2016,ivashenko_light_2016,imada_single-molecule_2017} and as a tool for molecular imaging with submolecular resolution.\cite{doppagne_vibronic_2017}
It was also employed for designing energetically efficient light emitting diodes.\cite{kimura_selective_2019}
Recently, strong light-matter interaction was measured in single molecule nanocavities. \cite{chikkaraddy_single-molecule_2016,benz_single-molecule_2016,chervy_vibro-polaritonic_2018,lombardi_pulsed_2018,chen_viewing_2010,reecht_electroluminescence_2014,ivashenko_light_2016}
While measurements in the strong coupling regime were performed in the absence of electron flux, extensions to current-carrying molecular junctions are expected soon.
Such developments will be an important step forward in the quest for optical control and characterization of molecular junctions.

Molecular spintronics is another rapidly growing field of research closely related to construction of nanodevices for quantum information. \cite{rossler_transport_2015} Here, the focus is on the flow of spin through the junction, rather than that of charge.
Measurements of spin polarized currents,\cite{petta_spin-dependent_2004,ham_spin_2012} 
spin-flip IETS,\cite{zyazin_electric_2010,ham_imaging_2013} 
spin interactions,\cite{czap_probing_2019,de_bruijckere_ground-state_2019,landig_microwave-cavity-detected_2019}
molecular spin selectivity,\cite{xie_spin_2011,naaman_chiral-induced_2012,burzuri_direct_2012,banerjee-ghosh_separation_2018,valli_quantum_2018} 
spin valves,\cite{urdampilleta_supramolecular_2011,urdampilleta_molecular_2011}
and photospintronic effects \cite{schreiber_coupling_2011,mondal_photospintronics:_2016} have all been reported in the literature. 

In strongly interacting junctions at low temperatures, spin exchange interactions can dress unpaired spins in the molecule with a cloud of opposite spins in the contacts.
The result is a correlated many-body electronic state comprising electrons in a physical region that can be much larger than that of the molecule.
This correlation reveals itself as a zero-bias conductance peak, known as the Kondo or Abrikosov--Suhl resonance, and has been observed in transport measurements in single atom \cite{park_coulomb_2002} and single molecule \cite{yu_kondo_2004,yu_transport_2004,yu_kondo_2005,scott_kondo_2010,scott_transport_2013}  transistors.
The behavior of the Kondo effect far from equilibrium (beyond linear response) is still not fully understood, but its signature has been observed in carbon nanotubes \cite{paaske_non-equilibrium_2006} and $\pi$-conjugated (OPV-5) molecules.\cite{osorio_electronic_2007}
Vibrational sidebands to the Kondo peak,\cite{yu_inelastic_2004,fernandez-torrente_vibrational_2008}
as well as an IETS--RIETS transition in the presence of a Kondo resonance,\cite{yu_inelastic_2004,rakhmilevitch_electron-vibration_2014,rakhmilevitch_vibration-mediated_2015}
have also been reported in the literature.

Finally, modern measurement techniques allow access not only to the electronic flux, but also to its noise or variance, and sometimes higher moments. \cite{djukic_shot_2006,kiguchi_highly_2008,tal_electron-vibration_2008,kumar_detection_2012,lumbroso_electronic_2018}
These quantities are related to the cumulants of the full counting statistics (FCS) generating function of electronic transport. \cite{levitov_charge_1993,levitov_electron_1996,blanter_shot_2000,esposito_nonequilibrium_2009}
Noise and higher moments can provide information not accessible via flux measurements, e.g. the number of transport channels in the junction and the effective charge of carriers. \cite{esposito_nonequilibrium_2009}
This paradigm has important implications: for example, noise measurements could potentially yield information on entanglement between degrees of freedom in the system\cite{klich_quantum_2009,song_bipartite_2012,gasse_observation_2013} or serve as an important ingredient in experimental validation for theories of nanoscale quantum thermodynamics.\cite{koski_distribution_2013} 

Today, molecular electronics is a thriving and interdisciplinary area of research involving researchers from fields as diverse as condensed matter physics and statistical mechanics; nonlinear optics, plasmonics and nanoscience; quantum chemistry and chemical dynamics; and a wide variety of engineering disciplines.
Single molecule junctions are used as testbeds for the study of fundamental physical properties of matter at nanoscale.
They are also widespread in more applied contexts, as models for quantum devices enabling energy transduction and storage, and as potential elements in proposed architectures for classical and quantum computers. \cite{joachim_electronics_2000,nitzan_electron_2001,van_der_molen_visions_2013,aradhya_single-molecule_2013,evers_preface:_2017}

Throughout the evolution of molecular electronics, the tremendous progress in experimental techniques has presented theorists with increasingly challenging and complex questions.
Molecular junctions are open quantum systems that can be driven far from their equilibrium state, and are characterized by a plethora of interactions at widely varying relative magnitudes.
This makes them attractive candidates for exploring a rich variety of fundamental quantum many-body physical problems in limits that remain poorly understood.
This physics is very different from better understood limits that can be treated by standard  approximations built around mean field arguments, which implies that rigorous theoretical treatments must rely on the development and implementation of specialized and advanced new methods.
Here, we present a short overview of recent developments in the Green's function techniques
for single molecule junctions.
Our consideration is mostly focused on method development.
We note that a complementary review focused on applications was recently published in Ref.~\onlinecite{thoss_perspective:_2018}.

The rest of our paper is structured as follows: in Section~\ref{overview}
we give a short overview of (non-Green's function based) theoretical methods utilized in transport problems.
Section~\ref{GFmethods} specifically focuses on Green's function methods for molecular electronics.
We discuss the standard nonequilibrium Green's function (NEGF) approach formulated around the noninteracting limit, then its many-body flavors, pseudoparticle (PP-) and Hubbard NEGF.
We then highlight recent developments in the superperturbation approach to transport, briefly review the state-of-the-art on time dependent problems, and go on to describe numerically exact quantum Monte Carlo approaches to accessing nonequilibrium Green's functions.
Section~\ref{conclude} summarizes the review.


\section{Overview of theoretical methods}\label{overview}
Before discussing theoretical methods, we begin by defining a concrete model of a molecular junction.
We consider a molecule $M$ coupled to a set of baths $B$. 
The molecule may also be subjected to external (classical) driving.
Baths usually encompass fermionic electronic contacts or leads, modeled as reservoirs of free charge carriers, each of which is taken to be in its own equilibrium state.
They can also include bosonic environments representing phonons and/or the optical modes of a quantized radiation field (see Fig.~\ref{gf_regimes}).
The Hamiltonian is taken to be
\begin{equation}
 \hat H(t) = \hat H_M(t) + \sum_B\bigg( \hat h_B + \hat V_{MB}(t)\bigg),
\end{equation}
where $\hat H_M(t)$ and $\hat h_B$ are the molecular and bath Hamiltonians, respectively. 
$\hat V_{MB}(t)$ describes coupling between the molecule and baths.
We note that the molecular Hamiltonian may describe several degrees of freedom (e.g., electronic and vibrational) and coupling between them, and can be interacting.
The baths are taken to be noninteracting, in the sense that their Hamiltonian is quadratic in each bath's set of fermionic/bosonic creation and annihilation operators.
The molecule--bath coupling $\hat V_{MB}(t)$ can also be interacting, for example when it describes electron--phonon coupling.
For most purposes in the field, it is sufficient to consider a bath Hamiltonian $\hat{h}_B$ that is both noninteracting (i.e. describes a normal metal or semiconductor) and time independent.
We note that generalization to include time-dependent driving in the baths (e.g. ac bias) is straightforward 
and for Green's function techniques follows the celebrated work by Jauho, Wingreen and Meir.\cite{jauho_time-dependent_1994}

A few simple cases are of special interest.
Below, if the molecular Hamiltonian $\hat{H}_M (t)$ and the coupling $\hat V_{MB}(t)$ are both quadratic, we refer to the entire system as noninteracting.
One canonical case that will be referred to below is the Anderson impurity model.
Here, only interacting term is of the form $U \hat{n}_{M\uparrow} \hat{n}_{M\downarrow}$, where $\hat{n}_{M\uparrow (\downarrow)}$ is a a spin up (down) population operator on the molecule; and there is no inter-spin coupling.
Another is the Holstein impurity model, where a single spinless electronic band is coupled to a single vibration or to 
a bath of phonons by a term of the form $\sum_{k}\sum_{\sigma\in\uparrow,\downarrow}\lambda_{\sigma}\hat{n}_{\sigma}\left(\hat{b}_{k}+\hat{b}_{k}^{\dagger}\right)$.
Here, $\hat{b}_{k}^{\dagger}$ ($\hat{b}_{k}$) creates (annihilates) a phonon in mode $k$.
If both interactions are present, the model may be called an Anderson--Holstein model.

Theoretical approaches to open quantum systems far from equilibrium differ in their treatment of both the system and bath degrees of freedom, but most rely either formally or in practice on time evolution from a convenient initial state to a more interesting, possibly stationary nonequilibrium state.
In particular, wavefunction methods consider the unitary time evolution generated by the Schr{\" o}dinger equation---complemented with appropriate boundary and initial conditions---in the entire system, comprising both the molecular and bath regions.
Density matrix approaches integrate out bath degrees of freedom, leading to a generally non-Markovian effective dynamics of the reduced density matrix as a function of time.
This encapsulates the effect of the baths into a modified nonunitary equation of motion known as a generalized quantum master equation.
Finally, Green's function methods consider the effective equations of motion of correlation functions rather than those of single-time properties.
At the very least, two-time correlations are used, and higher order, many-time correlations may also be evaluated and employed.
All three formulations are equivalent in the exact case, but most studies rely on approximations; here, these different languages lend themselves to very different schemes.
In what follows we will discuss the relative merits of Green's function approaches in the study of molecular junctions.
First, we briefly consider several important points regarding wavefunction and density matrix methods, without providing an exhaustive review.
Green's function approaches, our main focus, will be treated in more detail in Section~\ref{GFmethods}.

\subsection{One-body wavefunction methods}
In stationary transport problems wavefunction formulations are often combined with scattering theory.
Here, the set of scattering states are obtained by solving the Lippmann--Schwinger equation.
The solution is most often performed for an effective single-particle wavefunction.
This means that the many-body wavefunction is assumed to be a Slater determinant of single-particle orbitals, often referred to as a single reference state.
For noninteracting systems, where the many-body problem can be reduced to single particle description, such considerations are exact; in interacting systems, they become approximate.
For example, it is common to use the solutions of the Hartree--Fock or Kohn--Sham equations as effective single-particle orbitals.
The celebrated Landauer-B{\" u}ttiker formalism and its many extensions describes the application of noninteracting wavefunction methods to transport.
In one set of representative examples from the literature, a combination of density functional calculations for molecular electronic structure and wavefunction-based scattering theory was used in the simulation of elastic transport in junctions. \cite{lang_resistance_1995,lang_oscillatory_1998,di_ventra_first-principles_2000}

Single particle wavefunction methods, especially in the scattering state representation, are remarkably simple and computationally efficient.
Where the single particle picture remains accurate, they can be generalized in a straightforward manner to treating a large variety of interactions and observables.
These attractive properties have resulted in a massive popularity, and such methods are now ubiquitous.

For interacting systems in the Fermi liquid regime, at linear response from zero temperature equilibrium, an exact noninteracting picture is available that allows e.g. for the evaluation of currents and shot noise. \cite{sela_fractional_2006}
While a few other examples of this kind may exist, in general many-body interactions can be handled only approximately within the noninteracting framework.
For example, inelastic transport can be addressed by treating the electron--vibration coupling perturbatively within the Born approximation;\cite{chen_inelastic_2004} and inelastic current noise has also been treated within a similar set of considerations. \cite{chen_effect_2005,kim_inelastic_2014}
A method treating electron--vibration interactions exactly in single electron scattering problems was introduced in Refs.~\onlinecite{bonca_effect_1995,bonca_inelastic_1997} and later applied to the description of cooperative effects in inelastic tunneling. \cite{galperin_cooperative_2013}
Single particle scattering theory was also employed in the derivation of current-induced forces on molecular nuclei \cite{bode_scattering_2011,bode_current-induced_2012,thomas_scattering_2012}
and in the quantum thermodynamics context. \cite{ludovico_dynamical_2014,gaspard_scattering_2015,bruch_landauer-buttiker_2018}
The relationship between single particle scattering-matrix and nonequilibrium Green's function formalisms 
for noninteracting systems is explained in detail in Refs.~\onlinecite{arrachea_relation_2006,wang_relation_2009}.

Another set of approaches based to some degree on the ability to understand single-particle scattering states relies on an exact mapping of steady-state nonequilibrium
to an effective equilibrium state,\cite{hershfield_reformulation_1993} which was shown to be equivalent to the Zubarev's nonequilibrium statistical operator method. \cite{ness_nonequilibrium_2013}
This mapping was utilized in studies of transport in the resonant level,\cite{han_quantum_2006} Anderson impurity \cite{han_mapping_2007} and  
Anderson--Holstein \cite{han_nonequilibrium_2010} models.
An interesting development in this regard is that nonequilibrium steady state transport properties can be obtained from imaginary time quantities, albeit with the need for analytical continuation. \cite{han_imaginary-time_2007,han_imaginary-time_2010}

\subsection{Many-body wavefunction methods}
Where interactions are strong enough that the single-particle picture no longer holds, it is necessary to explicitly consider wavefunctions in the full many-body Hilbert space.
The size of this (multireference) space increases exponentially with the number of electrons in the system, which is always infinite in scenarios describing electronic transport across junctions.
Therefore, use of many-body wavefunctions necessarily requires efficient truncated representations of the Hilbert space in practice.
A variety of celebrated numerical wavefunction schemes have employed some variation of this idea to address transport problems, including the numerical renormalization group (NRG) method,\cite{anders_real-time_2005, anders_steady-state_2008} matrix product state (MPS) techniques,\cite{schmitteckert_nonequilibrium_2004} multiconfiguration Hartree--Fock (MCTDH) and its later,\cite{wang_numerically_2009, wang_numerically_2011, wang_numerically_2013} and more recent modifications. \cite{schwarz_lindblad-driven_2016, schwarz_nonequilibrium_2018}
We will very briefly go over some of these numerically exact approaches.

NRG relies on extracting a small, effective many-body Hamiltonian from the infinite interacting problem by way of a normalization procedure.
Nonequilibrium problems can be addressed directly in steady state by considering scattering states,\cite{anders_steady-state_2008} or by following the time evolution from some initial state. \cite{anders_real-time_2005}
Variations of this idea have been applied to steady state \cite{anders_quantum_2010,schmitt_comparison_2010} and transient properties \cite{anders_real-time_2005,anders_spin_2006,guttge_hybrid_2013} of strongly correlated junctions.

Applications of MPS techniques to transport began to appear in the literature before that term came to be widely used, initially in the form of time-dependent density matrix renormalization group (DMRG) calculations. \cite{white_real-time_2004,dias_da_silva_transport_2008,kirino_time-dependent_2008,guo_density_2009,kirino_nonlinear_2011}
Shot noise and FCS within the self-dual interacting resonant level model (an exactly solvable model for interacting transport) have also been evaluated.\cite{brandbyge_density-functional_2002,carr_full_2011,carr_full_2015}
In the language that has emerged over the years, these simulations rely on the MPS, an efficient and accurate ansatz for the ground state of one-dimensional systems, and understanding this has lead to concrete algorithmic improvements. \cite{schollwock_density-matrix_2005,mcculloch_density-matrix_2007,dorda_auxiliary_2015,werner_positive_2016}
It is possible to propagate the complete system ($M+B$) in time after an initial quench, such as the sudden application of a bias voltage; or to directly find an MPS representing a nonequilibrium steady state\cite{mascarenhas_matrix-product-operator_2015,cui_variational_2015}.
The main limitation of MPS methods is their inability to describe states with long-ranged entanglement along the chain.
This often makes it difficult to reach long simulation times, though this restriction can be greatly ameliorated by a good choice of basis.\cite{wolf_solving_2014,he_entanglement_2017,he_entanglement_2019,rams_breaking_2019,wojtowicz_open_2019,krumnow_towards_2019}
Higher dimensional systems also pose a challenge to MPS methods and other tensor network techniques, even in equilibrium.

Multilayer, multiconfiguration time-dependent Hartree (ML-MCTDH) theory 
employs a time-dependent multiconfigurational expansion of  wavefunctions that are propagated in time by using the Dirac--Frenkel variational princle.
The method, originally formulated in Ref.~\onlinecite{wang_multilayer_2003} for distinguishable particles, was extended to account for bosonic and fermionic statistics in Ref.~\onlinecite{wang_numerically_2009}.
The method has been applied to heat transport \cite{velizhanin_heat_2008} and electronic transport in junctions with electron--electron \cite{wang_multilayer_2018} and electron--phonon interactions.\cite{wang_numerically_2011,wang_numerically_2013,wilner_nonequilibrium_2014,wang_accuracy_2016}
In the latter case, the possibility of bistability in nonequilibrium quantum systems coupled to vibrational degrees of freedom\cite{galperin_hysteresis_2004} was explored.\cite{wilner_bistability_2013,wilner_phonon_2014}

\subsection{Density matrix formulations}
The density matrix $\rho$ of the full system, including both the molecular region $M$ and the baths $B$, evolves in time according to the Liouville--von-Neumann equation.
However, when the wavefunction is determined by $N$ coefficients, the density matrix is determined by $N^2$; and in a many-body system, $N$ is exponential in the physical system size.
This means that dealing with density matrices of the full system is generally more expensive than dealing with wavefunctions.
However, it is possible to consider the reduced density matrix $\sigma\equiv\textrm{Tr}_B\left\lbrace\rho\right\rbrace$, which has a dimensionality determined entirely by the size of the molecular region $M$.
We will broadly refer to methods formulated in terms of the reduced density matrix as ``density matrix methods''.

Diagonal elements of $\sigma$ yield information about probabilities to observe molecular states,
while off-diagonal elements relate to bath-induced coherences between such states.
Its exact dynamics is given by the Nakajima--Zwanzig--Mori generalized quantum master equation (GQME),\cite{zwanzig_nonequilibrium_2001} but are often treated perturbatively in $\hat{V}_{MB}$.\cite{montoya-castillo_approximate_2016,kelly_generalized_2016}
The lowest (second) order expansion in the molecule--bath coupling, along with a Markovian assumption, is known in the quantum transport context as the Born--Markov or Redfield approximation. \cite{breuer_theory_2003}
The latter is not adequate in representing bath-induced coherences,\cite{esposito_efficiency_2015,gao_simulation_2016}
and is problematic in the presence of degeneracies. \cite{schultz_quantum_2009}
The Redfield QME is often written within an additional rotating wave/secular approximation (RWA), where dynamics of the populations (or the Pauli equations) is decoupled from that of the coherences. \cite{breuer_theory_2003}
This has the advantage of guaranteeing positive definite reduced dynamics.
The Pauli equations are formulated in the molecular eigenbasis with rates given by the Fermi golden rule.

Markovian Pauli equations have been successfully employed in far too many contexts to provide an exhaustive list here.
One interesting example is the modeling of transport in donor--bridge--acceptor molecular complexes;\cite{muralidharan_probing_2006,migliore_nonlinear_2011,migliore_irreversibility_2013,craven_electron_2016,craven_electrothermal_2017} another is the description of inelastic transport.\cite{siddiqui_phonon_2007,rebentrost_environment-assisted_2009}
Many Redfield QME studies also go beyond the level of Pauli equations, providing some account of quantum coherence. \cite{harbola_quantum_2006,dubi_thermoelectric_2009,gelbwaser-klimovsky_strongly_2015}
Nevertheless, while attractive due to their simplicity, Markovian Redfield QMEs fail to account for molecule--bath correlations and hybridization, and therefore cannot be reliably employed in all regimes.

We note in passing that similar considerations employing bare perturbation theory in the molecule--bath coupling are at the heart of nonlinear optical spectroscopy studies. \cite{mukamel_principles_1995,harbola_nonequilibrium_2006,harbola_frequency-domain_2014,goswami_electroluminescence_2015,agarwalla_coherent_2015}
While completely justifiable in isolated systems or for strictly Markovian open system dynamics, utilization of bare perturbation theory in open systems
with essentially non-Markovian dynamics is known to violate conservation laws,\cite{baym_conservation_1961,baym_self-consistent_1962} and bare perturbative corrections can lead to qualitatively incorrect predictions. \cite{gao_optical_2016}
Recently, a conserving flavor of nonlinear optical spectroscopy methods based on GF analysis was presented in Ref.~\onlinecite{mukamel_flux-conserving_2019}.
We also note that treatments of dynamics involving a combination of non-Markovian and kinetic schemes have been employed in the literature.\cite{lu_laserlike_2011,simine_vibrational_2012,gelbwaser-klimovsky_high-voltage-assisted_2018,foti_origin_2018}
Such mixing requires care, as it may otherwise lead to qualitative failures in the analysis\cite{nitzan_kinetic_2018}.

Using the rigorous GQME as a theoretical framework, it is possible to perform expansions in the molecule--bath coupling 
$\hat{V}_{MB}$ to higher orders.
As with any perturbation theory, the semianalytical implementation becomes expensive for high-order self energies, but is extremely efficient and powerful when low orders are appropriate.
Rules for the construction of Feynman diagrams and their resummation or dressing, as well as an application of the resulting theory to transport processes, were presented in Ref.~\onlinecite{schoeller_mesoscopic_1994}.
This machinery was then successfully applied to the description of various transport problems. \cite{konig_zero-bias_1996,konig_cotunneling_1997,leijnse_pair_2009}
A simplification resulted in formulation of a kinetic equation approach to transport,\cite{leijnse_kinetic_2008} where rates were calculated within fourth order perturbation theory.
Similar fourth order considerations were employed in Ref.~\onlinecite{jang_fourth-order_2002},
and later developments allowed for a reduction in the number of diagrams. \cite{koller_density-operator_2010}
We note that the Liouville space formulation is complicated by the need to define diagrams with respect to ordering on both the Keldysh contour and the real time axis.
Green's function analysis, where ordering along the real axis can be dropped, allows some simplifications with respect to this. \cite{bergmann_electron_2019}

\subsection{Numerically exact approaches to the density matrix}
Most numerically exact methods and many approximate ones 
employ time propagation and are limited to short times, 
in the sense that the computational cost of simulating dynamics scales more than linearly with time.
Another application of the exact GQME leverages the rapid decay of the memory kernel in time that is characteristic to certain physical regimes. \cite{cohen_memory_2011}
If the memory kernel for the reduced dynamics of the molecular region $M$ can be simulated up to the time where it becomes negligible, the GQME can then be used to evaluate dynamics to any later time.
An extreme example of this is the Markovian limit, where evaluation of the kernel over an infinitesimal timescale suffices to fully describe the local population dynamics at any later time.
Evaluating the memory kernel is nontrivial, and one successful scheme, 
pioneered in Refs.~\onlinecite{shi_new_2003,zhang_nonequilibrium_2006} 
for the spin--boson model, relies on the evaluation of a set of 
auxiliary observables and the solution of an integral equation involving them.
This scheme was applied to transport through a junction with electron--electron interactions using real time Monte Carlo; \cite{cohen_memory_2011,cohen_numerically_2013} to inelastic vibrational transport using MCTDH;\cite{wilner_bistability_2013,wilner_nonequilibrium_2014} and as a way of enhancing semiclassical simulations that are more accurate at short timescales.\cite{kelly_efficient_2013,kelly_generalized_2016}
Alternative schemes for obtaining the memory kernel without the need for evaluating auxiliary observables later emerged,\cite{kidon_exact_2015,kidon_memory_2018} though with an added requirement for higher-order derivatives of the dynamics that can sometimes be numerically problematic.
A discrete version of these ideas can be expressed in terms of transfer tensors, or dynamical maps that have a matrix product form. \cite{cerrillo_non-markovian_2014}

Next, we mention two very successful numerically exact techniques that explicitly employ the density matrix picture in their construction: the hierarchical equation of motion (HEOM, sometimes also called the hierarchical quantum master equation or HQME in the literature), and the iterative summation of path integrals (ISPI).

The HEOM was originally introduced to address dissipative chemical dynamics,\cite{tanimura_time_1989,tanimura_quantum_2009} but was later generalized to models of quantum transport. \cite{jin_exact_2008,zheng_numerical_2009}
Beginning with an exact solution at the atomic limit, the idea is to write a systematic expansion in terms of a series of increasingly high-order auxiliary density matrices that contain information about mixed molecule--bath quantities.
This hierarchy can be efficiently evaluated to very high orders for certain forms of the lead density of states such as,
for example, a linear combination of several Lorentzian functions.
The expansion is eventually truncated at some finite order when a desired level of accuracy is reached.
In most cases, many of the auxiliary density matrices have little or no effect and can be dropped.

Early versions of HEOM were limited to high temperatures and specific band structures,\cite{hartle_decoherence_2013} but the approach has since undergone rapid progress. \cite{li_hierarchical_2012,hartle_decoherence_2013,ye_heom-quick:_2016,schinabeck_hierarchical_2018,erpenbeck_hierarchical_2019} 
Several works have considered time-dependent transport characteristics,\cite{hartle_formation_2014,cheng_time-dependent_2015,erpenbeck_hierarchical_2019} and HEOM has the important advantage over most numerically exact techniques that computational scaling in time is linear.
New developments enable trading some of this advantage for greatly improved access to more general band structures 
and lower temperatures.\cite{erpenbeck_extending_2018}
It should soon become clear to what degree these novel techniques enable access to correlated physics at low temperatures. 

The ISPI method,\cite{weiss_iterative_2008,segal_numerically_2010} like the HEOM, traces its roots to a predecessor in chemical dynamics. \cite{makarov_path_1994,makri_numerical_1995,makri_tensor_1995}
This method is based on expressing the action of the time evolution operator $e^{-i\hat{H}t}$ on an initial density matrix and measured operator as a path integral over a set of intermediate molecular states at a set of discrete times.
The contribution of each path, which is analogous to the Feynman--Vernon influence functional, is then truncated beyond a finite memory time.
Using this assumption, the ISPI method expresses time dependent observables in such a way that the dominant computational cost becomes exponential in the ratio between the memory time and the time discretization interval.
This becomes numerically exact for small time discretization when the truncation is applicable.

ISPI methods have been used to explore fermionic transport through quantum junctions in the presence of many-body interactions, in the Anderson impurity model and its spinless counterpart,\cite{segal_nonequilibrium_2011,becker_non-equilibrium_2012} explored interference effects in the presence of a magnetic field \cite{bedkihal_dynamics_2012-1} and employed to benchmark simulation methods. \cite{agarwalla_anderson_2017}
Systems simultaneously coupled to both fermionic and bosonic reservoirs were also considered. \cite{hutzen_iterative_2012,simine_path-integral_2013}
The method was recently generalized to allow for calculating the FCS of thermal transport in the spin--boson model. \cite{kilgour_path-integral_2019}

We note in passing that NRG and DMRG are also able to provide access to density matrices, though for most applications this is more expensive than access to wavefunctions.
In particular, the coupling of effective Markovian baths has been shown to be a promising route for obtaining accurate results at long dynamical timescales or steady state.\cite{dorda_auxiliary_2015,schwarz_lindblad-driven_2016,fugger_nonequilibrium_2018,sorantin_auxiliary_2019}

\subsection{Semiclassical approaches}
In an effort to describe molecules coupled to anharmonic environments, a semiclassical approach to quantum transport was proposed in Ref.~\onlinecite{swenson_application_2011}.
Classical dynamics can be simulated at polynomial complexity in both time and system size, whereas (exact) quantum dynamics always carries an exponential cost.
The main idea is therefore to enable simulation of quantum dynamics by mapping a second-quantized Hamiltonian onto a corresponding classical system.
The classical model is then solved using standard molecular dynamics techniques.

While the mapping suggested by Swenson et al. is approximate, the approach was shown to correctly reproduce the time dependent characteristics of the resonant level model for elastic \cite{swenson_application_2011} and inelastic \cite{swenson_semiclassical_2012,li_quasi-classical_2014} transport.
The method was also applied to transport through chaotic cavities. \cite{novaes_semiclassical_2012,berkolaiko_combinatorial_2013}
Interesting developments include a classical mapping for Hubbard operators,\cite{li_classical_2014} an exact mapping for fermions,\cite{montoya-castillo_exact_2018} and an accurate quasiclassical map that captures both noninteracting 
fermion dynamics and Coulomb blockade effects.\cite{levy_complete_2019}
These methods are relatively young in the field and appear very promising, but how they end up fitting into the puzzle remains to be seen.

\subsection{Renormalization group methods} 
Renormalization group methods can provide highly accurate results under certain considerations, especially when the physics is dominated by modes in a restricted energy range, such as those associated with the Kondo effect.
The NRG, mentioned in our earlier discussion of wavefunction methods, is a numerically exact method based on such ideas.
Other RG methods employed in transport simulations  
are the perturbative RG, functional RG and similarity RG techniques.
The perturbative renormalization group (RG) method starts from 
a perturbative expansion up to the leading order in a logarithmic correction. \cite{paaske_nonequilibrium_2004,rosch_kondo_2005,mitra_current-induced_2011,chung_nonequilibrium_2013}
The Keldysh formulation of functional RG (fRG) has also been widely applied 
to transport problems.\cite{xu_spintronic_2008,jakobs_nonequilibrium_2010,schmidt_transport_2010,metzner_functional_2012,laakso_functional_2014,wentzell_magnetoelectric_2016}
Limits of validity of these powerful approaches is of great interest 
to the theoretical transport community.
One interesting example of such investigations is 
Ref.~\onlinecite{laakso_functional_2014}, 
where it was found that different results are predicted by 
the equilibrium and nonequilibrium formulations of fRG.

Somewhat special among RG approaches is the flow equation method or similarity RG,\cite{wegner_flow-equations_1994,wegner_flow_2001,kehrein_flow_2006} where the renormalization flow is implemented by a sequence of infinitesimal unitary transformations that leads to a diagonalized Hamiltonian taking all energy scales into account.
This is in contrast to usual RG approaches, where the flow leads to the truncation of higher energies; however, the method remains perturbative by nature, and cannot be applied at all parameter regimes.
The flow equation method has been successfully applied to the transient and steady state behavior of the nonequilibrium 
Kondo,\cite{kehrein_scaling_2005,fritsch_non-equilibrium_2009} spin--boson,\cite{hackl_real_2008,hackl_unitary_2009} resonant level \cite{wang_exact_2010} and Anderson impurity \cite{wang_flow_2010} models. 

The GQME-based perturbation theory has also been reframed in terms of a real time renormalization group method. \cite{schoeller_introduction_2000,pletyukhov_nonequilibrium_2012}
Real time renormalization group in the framework of kinetic equation was presented in Refs.~\onlinecite{saptsov_fermionic_2012,saptsov_time-dependent_2014}.

\subsection{Comparative work} 
Some of the approaches we have discussed are clearly suitable for certain tasks, and analytical considerations can often guide us in our choice.
Nevertheless, it is often difficult to understand which method should be used or what reliable benchmarks are available for a particular model and parameter regime, even in surprisingly simple cases.
Several studies have taken on the important work of comparing and sometimes contrasting the different approaches described in this section, as well as some of the Green's function approaches described below. \cite{eckel_comparative_2010,eckstein_new_2010,andergassen_charge_2010,koerting_non-equilibrium_2011,hamad_scaling_2015,hartle_transport_2015,nghiem_ohmic_2016,reimer_five_2019} 


\section{Green's function methods in molecular electronics}\label{GFmethods}
We now discuss the application of Green's function methods to transport through molecular junctions.
First, we consider the standard diagrammatic technique and its extensions.
This is followed by review of numerically exact approaches based on Green's functions.

\begin{figure}
\centering\includegraphics[width=\linewidth]{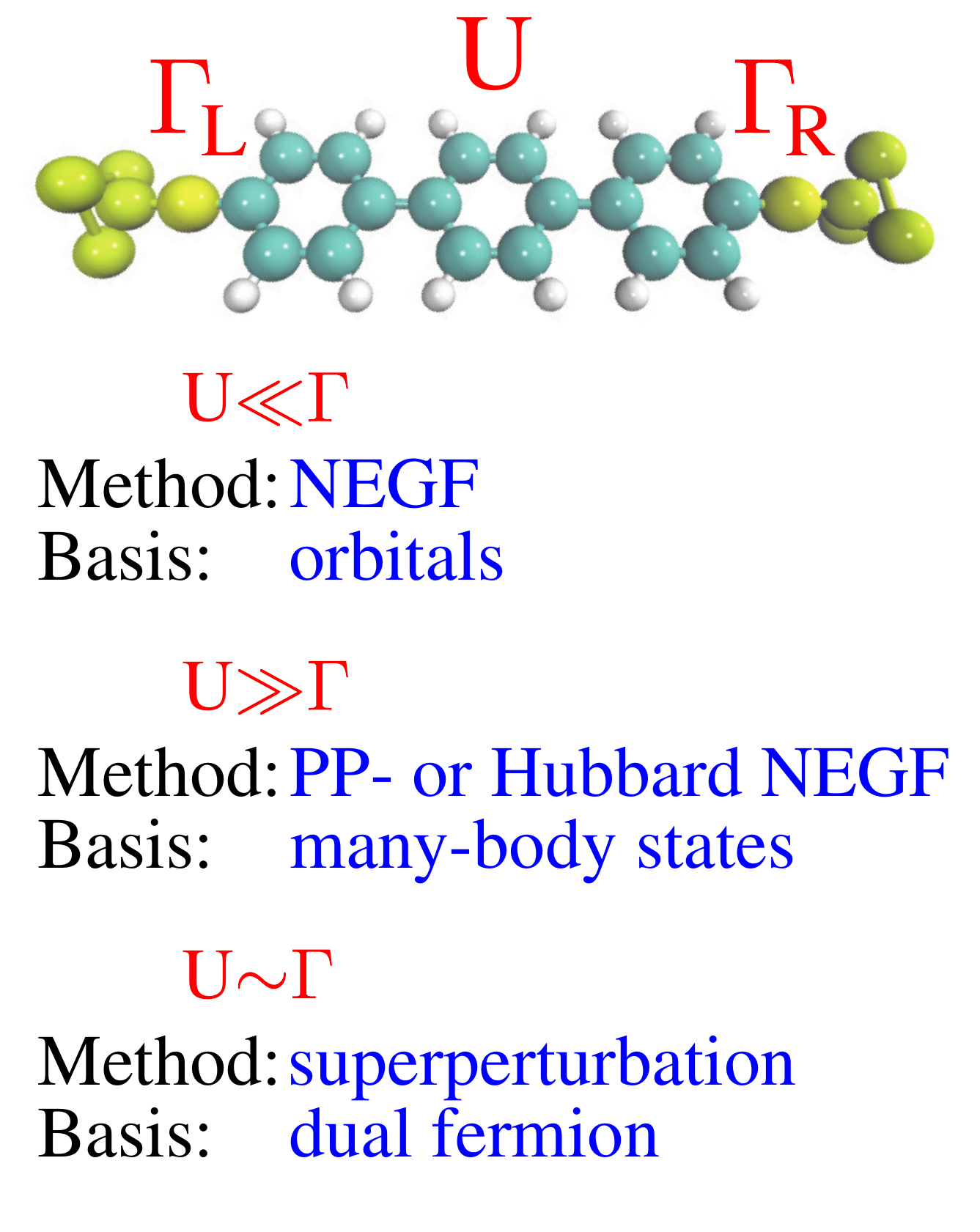}
\caption{\label{gf_regimes}
Convenient Green's function methods for different parameter regimes.
}
\end{figure}

In quantum many-body theory, the term ``Green's function'' (GF) actually refers to correlation functions between sets of two or more spatial locations or orbitals, taken at different times. The simplest one-body, or two-time GF takes on the following form:
\begin{equation}
\label{GFNEGF}
G_{m_1m_2}\left(\tau_1,\tau_2\right) = -i\langle T_c\,\hat d_{m_1}(\tau_1)\,\hat d_{m_2}^\dagger(\tau_2)\rangle.
\end{equation}
Here, $\hat d_m^\dagger$ ($\hat d_m$) creates (annihilates) an electron on molecular orbital $m$, $T_c$ is the contour ordering operator and the $\tau_{1,2}$ are times on the Keldysh contour. \cite{stefanucci_nonequilibrium_2013}

These objects are of interest for three main reasons.
First, empirically, \emph{few-body (few-time) GFs tend to correspond the kinds of observables that are easier to measure in experiments}.
For example, particle density, energy density and their fluxes can all be expressed via the simplest GFs, which have only two times.
These properties are easily measured in some very complex systems, whereas the many-body wavefunction is generally experimentally inaccessible.

Second, in all but the smallest systems, even if the wavefunction could be measured in complete detail, there is not enough data storage in the world to contain even a rough discretized description of it.
The amount of information in a wavefunction increases exponentially with the number of degrees of freedom.
On the other hand, \emph{GFs contain an amount of information that scales only polynomially with the system size}, with two-time GFs scaling quadratically.
In the noninteracting case efficiently expressed wavefunctions, density matrices and GFs all provide essentially equivalent information at polynomial scaling with the system size and simulation time.
Choosing between them is therefore often a purely aesthetic matter.
In the presence of many-body correlations, however, the latter two languages allow us to concentrate on a small subset of relevant degrees of freedom while avoiding any direct reference or access to the exponential Hilbert space of the leads.
While density matrix methods remain exponential in the dimension of the region of interest, GFs of any given order are described by correlation functions, the number of which is always polynomial.

Third and finally, GF theory embodies \emph{a powerful methodological framework for constructing approximation schemes}.
Concepts like resummation and conserving approximations make GFs extremely useful.
GFs also naturally allow for consistently and seamlessly treating some regions (say, the leads) as noninteracting or analytically solvable, while numerically accounting for detailed structure and interactions in other regions.

The nonequilibrium Green's function (NEGF) technique, which is built around (actually or effectively) noninteracting single-particle orbitals, has long been a tool of choice for ab initio simulations in molecular electronics.
The noninteracting NEGF is a convenient starting point for more accurate or general GF methods.
This requires identifying a small expansion parameter.
Two common choices are related to two generic energy scales that exist in essentially all single molecule junctions (see Fig.~\ref{gf_regimes} for an illustration).
The first is the strength of many-body interactions $U$: 
for example, intra-molecular electron--electron attraction, 
or coupling between electronic and vibrational degrees of freedom.
The second, $\Gamma$, is the coupling strength, or hybridization or escape rate, 
between the molecule and electronic leads.

When many-body interactions are much smaller than the hybridization, standard NEGF theory exactly solves for the GFs of the noninteracting problem modeled by $\hat{H}^{(0)}=\hat{h}_M+\hat{h}_B+\hat{v}_{MB}$.
Here, $\hat{h}_M$ and $\hat{v}_{MB}$ are the noninteracting (quadratic) parts of $\hat{H}_M$ and $\hat{V}_{MB}$, respectively.
The remainder, $\hat{V}=\hat{H}_M-\hat{h}_M+\hat{V}_{MB}-\hat{v}_{MB}$, becomes a small parameter, and perturbation theory in $\hat{V}$ can be efficiently employed.
The approaches we previously mentioned all rely on this idea.

The opposite limit can be addressed by exactly solving the leads and molecule separately: $\hat{H}^{(0)}=\hat{H}_M+\hat{h}_B$.
Here, the molecule at the so-called ``atomic limit'' is assumed to be tractable because it is a physically small region, comprising only a few interacting orbitals.
The hybridization, $\hat{V}=\hat{V}_{MB}$, is then treated perturbatively.
The pseudoparticle and Hubbard NEGF techniques to be discussed below are based on a hybridization expansion.

In the absence of a small parameter, it may be possible to take advantage of hybrid schemes like the superperturbation theory or dual fermion technique, to be discussed below.
The conditions for the applicability of such methods are more involved and not yet fully understood.
If approximate schemes cannot be established as accurate, one must rely on substantially more expensive numerically exact schemes, to which we refer 
at the end of the Section.
Below we discuss the various approximations with regard to applications 
in molecular electronics.


\subsection{Nonequilibrium Green's functions (NEGF)}
The NEGF method is an established and widely used technique
that is exact and efficiently solvable at the noninteracting limit, but also includes well understood machinery for taking into account weak many-body interactions within diagrammatic perturbation theory.\cite{danielewicz_quantum_1984,rammer_quantum_1986,wagner_expansions_1991,haug_quantum_2008,kamenev_field_2011,stefanucci_nonequilibrium_2013}
It relies on the fact that the GF of Eq.~\eqref{GFNEGF} obeys a Dyson equation:
\begin{equation}
\begin{aligned}G_{m_{1}m_{2}}\left(\tau_{1},\tau_{2}\right) & =G_{m_{1}m_{2}}^{\left(0\right)}\left(\tau_{1},\tau_{2}\right)\\
& +\sum_{m^{\prime}m^{\prime\prime}}\int_{C}\mathrm{d}\tau^{\prime}\mathrm{d}\tau^{\prime\prime}G_{m_{1}m^{\prime}}^{\left(0\right)}\left(\tau_{1},\tau^{\prime}\right)\\
& \Sigma_{m^{\prime}m^{\prime\prime}}\left(\tau^{\prime},\tau^{\prime\prime}\right)G_{m^{\prime\prime}m_{2}}\left(\tau^{\prime\prime},\tau_{2}\right).
\end{aligned}
\end{equation}
First, the noninteracting GF, $G_{m_{1}m_{2}}^{\left(0\right)}\left(\tau_{1},\tau_{2}\right)$, is obtained in the absence of many-body interactions.
One immediate advantage of the GF methodology is that it is possible to obtain all GF elements in the molecular region $M$ without solving for the dynamics of the full system.
The leads can be handled separately, and their GFs are usually solved for in the continuum limit.
The effect of the leads then enters as a self-energy term $\Sigma_{m^{\prime}m^{\prime\prime}}^{\mathrm{hyb}}\left(\tau^{\prime},\tau^{\prime\prime}\right)$ in the Dyson equation for the molecular GFs, and the effective dimension of the GF becomes the number of orbitals in the molecular region.

Second, interactions are taken into account diagrammatically, also taking the form of a set of many-body or interaction self-energy terms, $\Sigma_{m^{\prime}m^{\prime\prime}}^{\mathrm{int}}\left(\tau^{\prime},\tau^{\prime\prime}\right)$.
Self-energies are usually functionals of the GF, such that the solution of the Dyson equation often involves finding the fixed point of a self-consistency relation.
These self-energies are by definition zero in all noninteracting regions.

Given the GF of the junction, responses to external perturbations (fluxes, noises and higher cumulants) are obtained from, e.g., the celebrated Jauho--Wingreen--Meir formula \cite{jauho_time-dependent_1994} and related  expressions. \cite{park_self-consistent_2011}

In ab initio simulations, an effective noninteracting model of the molecular region is usually constructed from the Kohn--Sham orbitals of a density functional theory (DFT) simulation. 
DFT, widely utilized for electronic structure simulations, naturally combines
with NEGF, because both methods are formulated in the language of quasiparticles or single-particle orbitals.
Indeed, the combination of these methods, known as NEGF-DFT, is frequently utilized in the theoretical description of molecular electronics. \cite{xue_first-principles_2002,brandbyge_density-functional_2002}
For example, NEGF-DFT was successfully applied to studies of elastic transport in noninteracting systems,\cite{crljen_nonlinear_2005,wohlthat_ab_2007,hybertsen_amine-linked_2008,stokbro_first-principles_2008,cheng_situ_2011,vazquez_probing_2012}
as well as inelastic transport mostly in the off-resonant tunneling regime. \cite{van_leeuwen_first-principles_2004,van_leeuwen_first-principles_2004,sergueev_ab_2005,frederiksen_inelastic_2007,sergueev_inelastic_2007,kaasbjerg_charge-carrier-induced_2013}
We note in passing that approximate NEGF based theoretical schemes for treating resonant IETS were also suggested 
in the literature.\cite{galperin_resonant_2006,seoane_souto_dressed_2014}
The method was further employed in studies of energy transport,\cite{lee_heat_2013,zotti_heat_2014,cui_quantized_2017,cui_peltier_2018}
for evaluating shot noise,\cite{avriller_inelastic_2012}
and for studies of disorder in nonequilibirum systems.\cite{karlsson_disorder_2018}
Besides global responses, it provides access to local characteristics: bond \cite{solomon_exploring_2010,herrmann_designing_2011,solomon_when_2011} 
and current density \cite{walz_current_2014,walz_local_2015,wilhelm_ab_2015,cabra_simulation_2018} fluxes,
as well as local noise spectroscopy,\cite{cabra_local-noise_2018} have all been described within NEGF-DFT.

Notably, the use of the Kohn--Sham Hamiltonian as an effective noninteracting description does not have firm theoretical justifications; nevertheless, it works well in many cases.
It has been suggested in the literature that in certain nonequilibrium situations the essentially single-particle nature of this approach may lead to an incomplete description of transport. \cite{baratz_gate-induced_2013,baratz_correction_2013}
Finally, we remark that while static DFT is most often used, time-dependent DFT has also been employed within NEGF calculations. \cite{van_leeuwen_causality_1998,van_leeuwen_nonequilibrium_2003,galperin_linear_2008,hopjan_merging_2016}

When a molecule is coupled to metallic electrodes to form a junction, its frontier molecular orbitals are renormalized---i.e., shifted and broadened---by the proximity of the substrate.
This is due to a combination of delocalization of electrons across the junction and screened polar interactions between electrons in the different regions.
These effects are not accounted for in any simulations of the isolated molecule, or by DFT simulations of the full system.
To account for such effects, first principle simulations within the GW approximation \cite{hedin_new_1965,aryasetiawan_gw_1998} were proposed.\cite{neaton_renormalization_2006,van_setten_gw-method_2013,liu_accelerating_2019}
The GW approximation is a self-consistent GF approximation in principle, 
but most commonly used implementations employ one iteration of GW 
to make corrections to the quasiparticle orbitals obtained from DFT.
The combination of DFT, GW and NEGF was successfully used in a number of studies. \cite{thygesen_conserving_2008,strange_self-consistent_2011,strange_towards_2011,pedersen_quantum_2014,capozzi_single-molecule_2015,strange_interference_2015,refaely-abramson_origins_2017,refaely-abramson_first-principles_2019}
Nevertheless, the mean-field nature of GW and the lack of complete screening in junctions may lead to failure of the method when applied to transport simulations. \cite{spataru_gw_2009}

GW formulations remain very expensive.
We briefly note here two promising advances that have not yet been used in the context of transport, but would be of interest in reducing the numerical cost: 
first, range-separated hybrid density functionals can be tuned to provide band gaps on par with the accuracy of GW. \cite{refaely-abramson_quasiparticle_2012,egger_reliable_2015}
Second, stochastic formulations of GW allow simulations including thousands of electrons. \cite{neuhauser_breaking_2014} 

Significant effort has been devoted to developing NEGF-based theoretical methods for understanding optical measurements in current-carrying molecular junctions. \cite{galperin_photonics_2017}
For example, NEGF methods for linear optical response,\cite{galperin_optical_2006,galperin_optical_2006} 
current-induced light,\cite{galperin_current-induced_2005} Raman spectroscopy \cite{galperin_raman_2009,galperin_raman_2011,park_charge_2011,mirjani_probing_2012} and multi-dimensional spectroscopy \cite{gao_simulation_2016} have been formulated.
We note that traditional approaches to nonlinear optical spectroscopy,\cite{mukamel_principles_1995}
which employ bare perturbation theory, do not satisfy conditions for building conserving approximations in the NEGF diagrammatic approach. \cite{gao_optical_2016}
Recently, a flux-conserving version of nonlinear optical spectroscopy for application in open molecular systems was suggested. \cite{mukamel_flux-conserving_2019}
Note also that within diagrammatic expansions, one can only treat relatively weak light--matter interactions, when both light and matter degrees of freedom are treated quantum mechanically.
On the other hand, there is no such restriction on the strength of interaction with a classical radiation field.
In the latter case, the Maxwell--NEGF method was utilized for simulation of junctions driven by external classical fields. \cite{sukharev_transport_2010,white_molecular_2012}

NEGF has also often been applied to molecular spintronics.
A few examples are studies of magnetic field effects on electronic conduction,\cite{rai_magnetic_2012}
spin pumps,\cite{sun_spin_2003,fransson_inelastic_2010,jahn_organic_2013}
spin-flip IETS,\cite{fransson_spin_2009,fransson_theory_2010,hurley_spin-flip_2011,rai_spin_2012}
transport in helical molecules (e.g., DNA),\cite{sun_quantum_2005,guo_spin-selective_2012,guo_sequence-dependent_2012,guo_enhanced_2012,rai_electrically_2013,gersten_induced_2013,guo_spin-dependent_2014}
quantum interference in spintronic devices,\cite{saraiva-souza_molecular_2014}
thermospintronics. \cite{dubi_thermospin_2009,fransson_spin_2011,tang_thermodynamics_2018} 
Green's function methods recently also were employed in studies of formation and dynamics of skyrmions. \cite{lounis_magnetic_2012,psaroudaki_quantum_2017,yudin_light-induced_2017,bostrom_steering_2019}

\begin{figure}[htbp]
\centering\includegraphics[width=\linewidth]{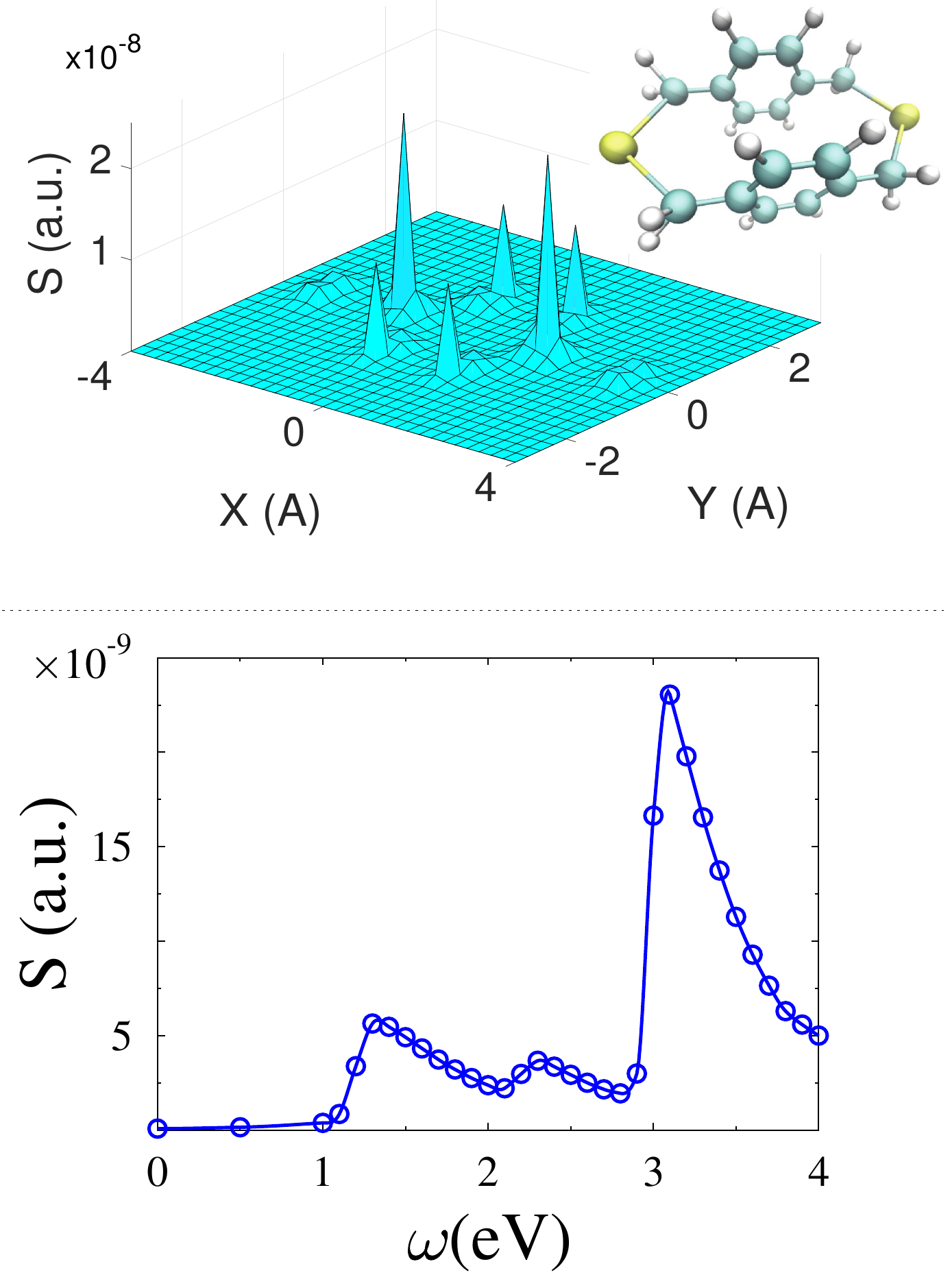}
\caption{\label{fig2}
Local noise cross-correlation as a probe of intra-system interactions in a molecular junction, illustrated in the top right corner.
(top) The cross-correlation map is shown in the XY plane at frequency $\omega=3.5$~eV.
(bottom ) the cross-correlation is plotted at the positions of the carbon atoms.
Reprinted with permission from G.~Gabra, M.~Di~Ventra and M.~Galperin, Phys. Rev. B \textbf{98}, 235432 (2018).
Copyright (2018) by the American Physical Society.
}
\end{figure}

NEGF has further been applied to study shot noise.
For example, shot noise noise spectroscopy of spin current \cite{wang_shot_2004,souza_spin-polarized_2008} and within inelastic transport \cite{zhu_theory_2003,galperin_inelastic_2006,haupt_phonon-assisted_2009,avriller_electron-phonon_2009,haupt_current_2010,urban_nonlinear_2010,novotny_nonequilibrium_2011} have been discussed in the literature.
The theory of light emission as a probe of electron shot-noise was put forward,\cite{kaasbjerg_theory_2015} and current fluctuations in the transient regime \cite{feng_current_2008,ochoa_pump-probe_2015} have been discussed.
The notion of local noise spectroscopy (LNS) (contrary to total junction noise) has also been introduced. \cite{cabra_local-noise_2018}
This concept is useful in theoretical consideration of, e.g, local molecular light emission patterns, heating maps, or mapping out local interaction within the junction (see Fig.~\ref{fig2}).
While in single molecule junctions the technique has purely theoretical applications, in extended systems (such as graphene nanoribbons) LNS predictions are measurable.
In addition to noise, higher moments and FCS have been considered.
As in flux modeling, study of FCS and waiting time distribution (WTD) for systems with strong many-body interactions is challenging within NEGF. \cite{gogolin_towards_2006,schmidt_charge_2009,maier_charge_2011,seoane_souto_transient_2015,tang_full-counting_2017}
At the same time, rigorous treatment of FCS within NEGF for noninteracting systems is 
well established in both the steady-state \cite{komnik_full_2014,esposito_efficiency_2015} and transient \cite{tang_waiting_2014,tang_full-counting_2014,yu_full-counting_2016} regimes.  

In summary, NEGF has many important advantages when applied to simulation of responses in single molecule junctions.
It has a well-established route to account for interactions within controlled diagrammatic perturbation
theory.
The conserving character of the resulting approximations \cite{baym_conservation_1961,baym_self-consistent_1962} is assured by utilization of the Luttinger--Ward functionals as generating functions for the self energies.\cite{luttinger_ground-state_1960,stefanucci_nonequilibrium_2013}
In the case of bilinear molecule--contacts coupling, the embedding or hybridization can be  exactly taken into account (i.e. the corresponding term in the self-energy is exact).
NEGF-DFT allows for convenient pairing between ab initio electronic structure calculations and the evaluation of transport properties.
FCS can be evaluated exactly for noninteracting and approximately for interacting systems, allowing for modeling of noise measurements in junctions.
Nevertheless, within the NEGF diagrammatic formulation it is challenging to treat strong many-body interactions.\cite{martin-rodero_interpolative_2008,von_friesen_successes_2009,schlunzen_dynamics_2016}
Also, information regarding molecular many-body states (as opposed to single-body orbitals) is more difficult to access, as it requires higher-order GFs.
Strong many-body interactions and state-specific information are of value for, e.g., the characterization of optoelectronic devices, where in order to measure optical response the molecule should be relatively weakly coupled to contacts.
A more convenient description of this regime is given by diagrammatic expansions formulated directly in terms of the interacting molecular many-body states.
Below we discuss GF techniques conveniently applicable in this regime.


\subsection{Pseudoparticle NEGF\label{ssec_pp_negf}}
Pseudoparticle (or auxiliary-operator) NEGF was first utilized in simulations of transport within the Anderson impurity model in Refs.~\onlinecite{meir_low-temperature_1993,wingreen_anderson_1994,sivan_single-impurity_1996,hettler_nonequilibrium_1998}.
More recently, such techniques have become useful in the context of driven materials, as solvers for the effective impurity models appearing in nonequilibrium dynamical mean field theory. \cite{aoki_nonequilibrium_2014}
We note in passing that the dynamical mean field theory, originally formulated for bulk materials, can also be employed in the treatment of molecular electronics including large interacting regions. \cite{jacob_dynamical_2010,turkowski_dynamical_2012}

The relationship between PP-NEGF and standard NEGF is established via a spectral decomposition of electron creation (annihilation) operators
$\hat d_m^\dagger$ ($\hat d_m$) in the many-body state basis $\{\lvert S\rangle\}$ of the isolated molecular region $M$.
Here, it is assumed to be a complete local basis:
\begin{equation}
\begin{aligned}\hat{d}_{m}^{\dagger} & =\sum_{S_{1},S_{2}}\left|S_{2}\right\rangle \left\langle S_{2}\right|\hat{d}_{m}^{\dagger}\left|S_{1}\right\rangle \left\langle S_{1}\right|\\
& \equiv\sum_{S_{1},S_{2}}\xi_{S_{2}S_{1}}^{m}\left|S_{2}\right\rangle \left\langle S_{1}\right|.
\end{aligned}
\label{spectral}
\end{equation}
PP-NEGF introduces second quantization in an extended Hilbert space, where pseudoparticle operators
$\hat p_S^\dagger$ ($\hat p_S$) create (destroy) many-body state $\lvert S\rangle$ starting from an unphysical vacuum state $\lvert \mathrm{vac}\rangle$
\begin{equation}
\label{PP2nd}
\lvert S\rangle=\hat p_S^\dagger \lvert \mathrm{vac}\rangle
\end{equation}
Similarly to the NEGF, Eq.~\eqref{GFNEGF}, the central object of interest is the single pseudoparticle GF
\begin{equation}
\label{PPNEGF}
 G_{S_1S_2}(\tau_1,\tau_2) = -i\langle T_c\,\hat p_{S_1}(\tau_1)\,\hat p_{S_2}^\dagger(\tau_2) \rangle.
\end{equation}

The formulation of diagrammatic technique in the PP-NEGF follows closely that of standard NEGF, with the difference that the NEGF expansion in weak many-body interaction is substituted by a PP-NEGF expansion in weak system--bath coupling.
The PP-NEGF therefore expands around the nonequilibrium atomic limit.
A more important difference is that the PP-NEGF is formulated in an extended Hilbert space, the physical subspace of which is defined by the normalization condition
\begin{equation}
\label{PPnorm}
\sum_S \hat p_S^\dagger\hat p_S =1.
\end{equation}
The necessity to impose the normalization condition on unrestricted solution is the weakest point in the PP-NEGF methodology.
Contrary to standard NEGF, where GFs are correlation functions between excitation operators, PP-NEGF considers correlations between states themselves.
In this sense it is closer in spirit to QME formulations, but allows for a more natural treatment of non-Markovian effects.

\begin{figure}[t]
\centering\includegraphics[width=\linewidth]{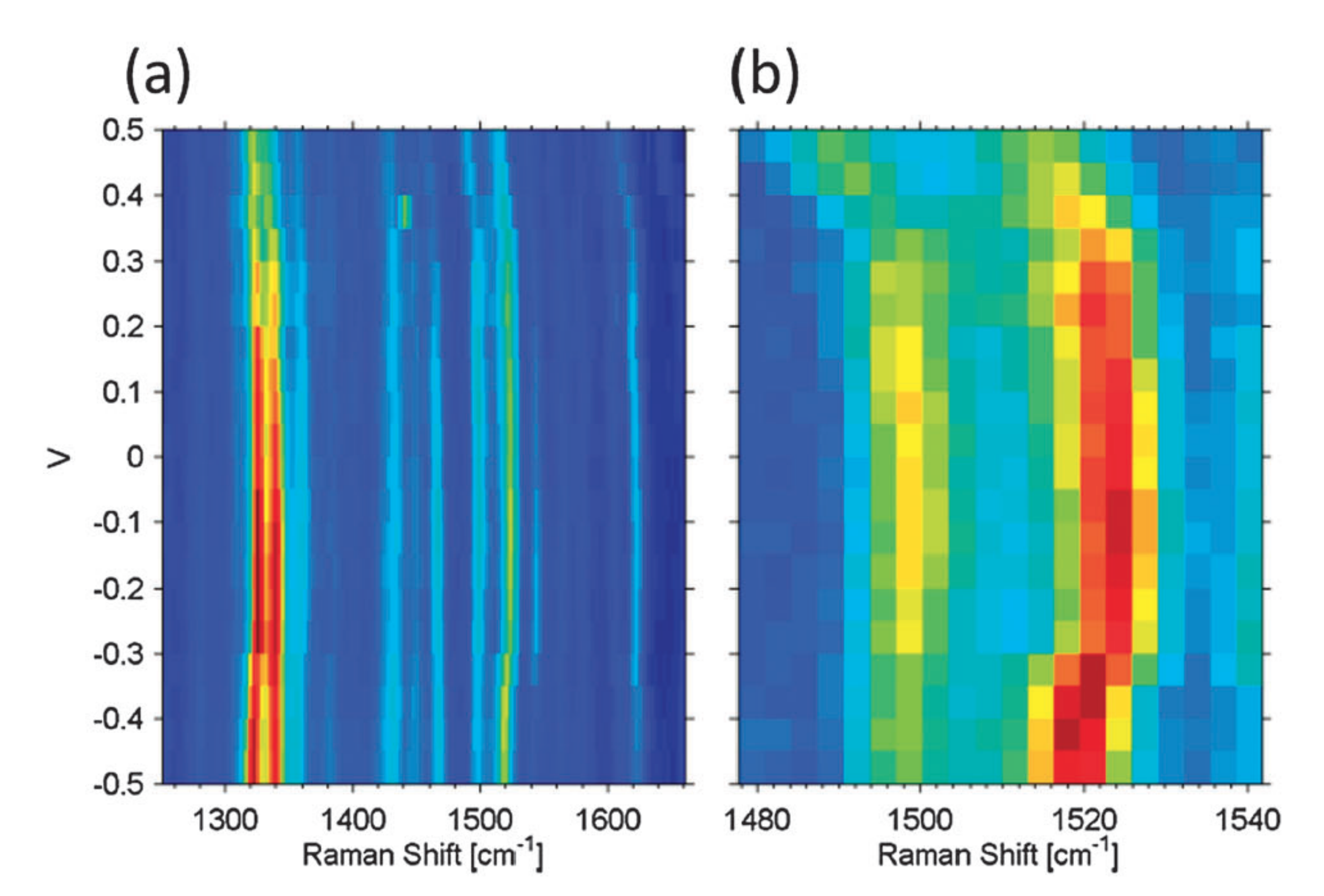}\\
\centering\includegraphics[width=\linewidth]{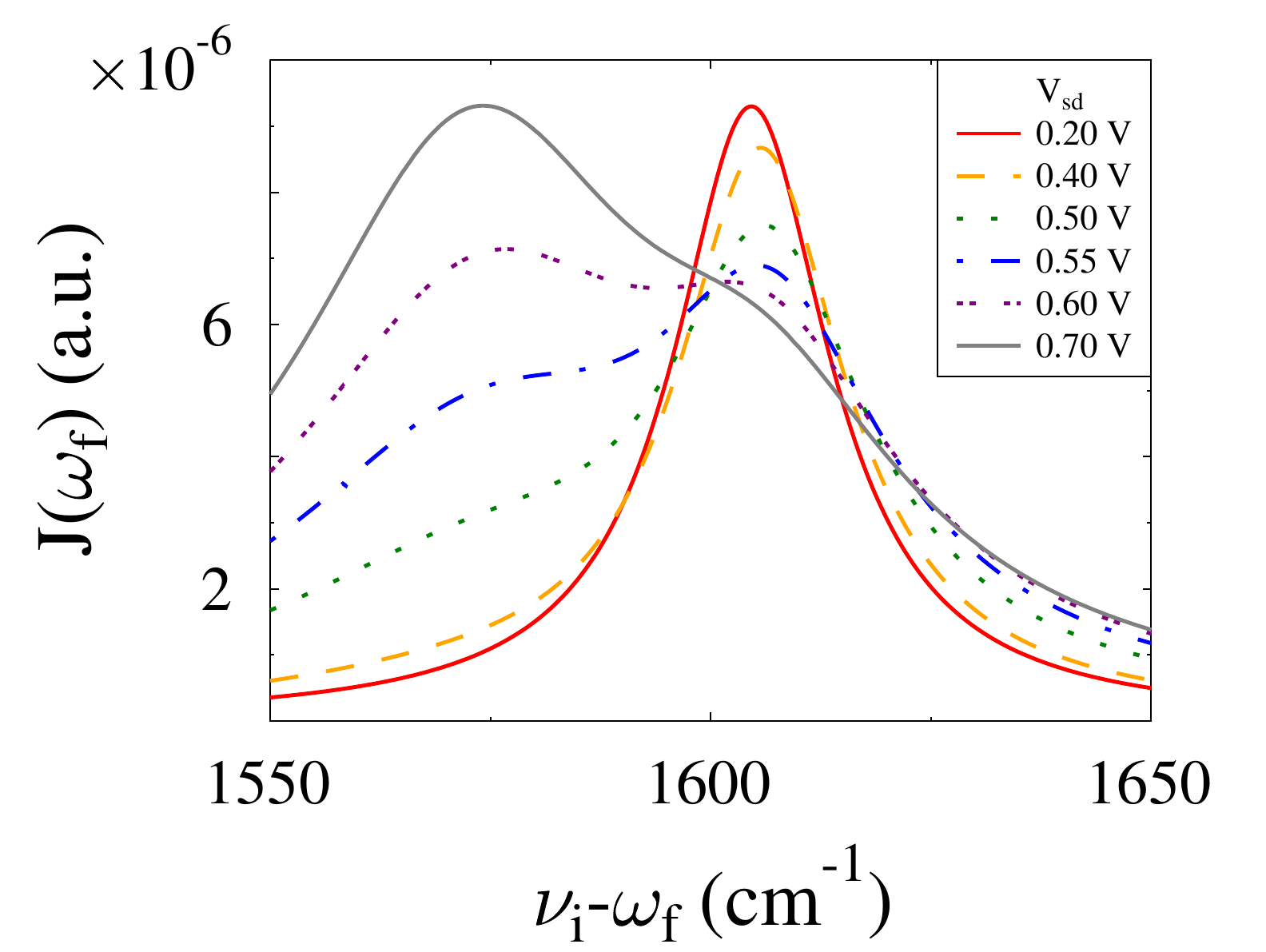}
\caption{\label{fig3}
Stokes shift due to charging in oligophenylene vinylene molecular junction.
Shown are experimental results (top) and PP-NEGF simulation of Raman scattering (bottom).
Top figure is republished with permission of Royal Society of Chemistry, from 
{\em Nanogap structures: combining enhanced Raman spectroscopy and electronic transport}, 
D.~Natelson, Y.~Li, and J.~B.~Herzog, Phys. Chem. Chem. Phys. \textbf{15}, 5262-5275 (2013); 
permission conveyed through Copyright Clearance Center, Inc.
Bottom figure is reprinted (adapted) with permission from  
A.~J.~White, S.~Tretiak, and M.~Galperin, Nano Lett. \textbf{14}, 699-703 (2014).
Copyright (2014) American Chemical Society.
}
\end{figure}

PP-NEGF was shown to be useful in studying transport through strongly correlated systems,\cite{eckstein_nonequilibrium_2010,oh_transport_2011} in modeling inelastic transport in the resonant tunneling regime with strong electron--vibration interactions or in presence of anharmonicities,\cite{white_inelastic_2012}
in treating electronic and energy transport on the same footing,\cite{white_collective_2012,white_coherence_2013}
and in first principles simulations of Raman spectroscopy in single molecule junctions \cite{white_raman_2014} (see Fig.~\ref{fig3}). 

The ability to access state-specific information also makes PP-NEGF useful in understanding nonadiabatic molecular dynamics in the regime of weak but non-vanishing molecule--contact electron exchange coupling.
In particular, in Ref.~\onlinecite{galperin_nuclear_2015} PP-NEGF was utilized to derive equations for the nuclear dynamics that reproduce the surface-hopping formulation in the limit of small coupling, and Ehrenfest dynamics when information on different charging states of the molecule is traced out.

We note that closely related to PP-NEGF methods are those where a reduced molecular propagator between contour times, $\mathrm{Tr}_{B}\left\{ \rho_{B}e^{-i\int_{z_1}^{z_2}\hat{H(z)}}\right\}$, replaces the GF as the fundamental object.
Expansions of this type appear as initial approximations in numerically exact schemes. \cite{gull_bold-line_2010,gull_numerically_2011,cohen_numerically_2013,cohen_greens_2014}
Propagator formulations trade some of the useful machinery of GF theory for the ability to treat intra-molecular interactions exactly while expanding in the molecule--bath hybridization, without the need for artificially extending the Hilbert space as in PP-NEGF.
Propagator methods do enable a type of modified diagrammatic resummation, and can be used to construct conserving approximations.
For instance, they have been used to explore transport in junctions with both electron--electron and electron--phonon interactions within the Anderson--Holstein model,\cite{chen_anderson-holstein_2016} as well as in a study (with comparisons to exact results) of the splitting of the Kondo conductance peak resulting from a nonequilibrium bias voltage. \cite{krivenko_dynamics_2019}

\begin{figure}[t]
\centering\includegraphics[width=0.9\linewidth]{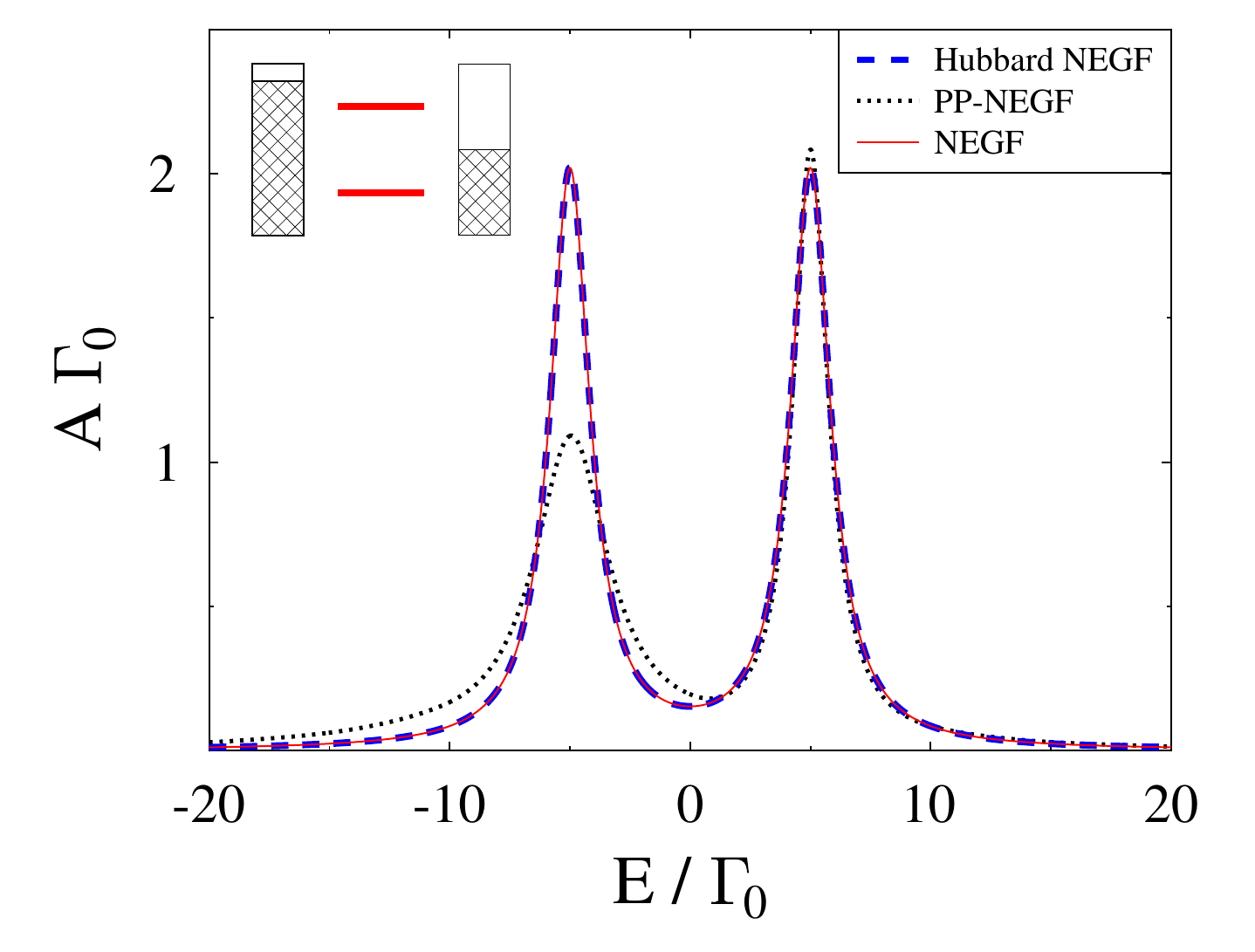}
\caption{\label{fig4}
Spectral function for noninteracting two-level system (inset).
Shown are (exact here) NEGF (solid line, red); PP-NEGF (dotted line, black); and Hubbard NEGF (dashed line, blue) results.
Reprinted from K.~Miwa, F.~Chen, and M.~Galperin, Sci. Rep. \textbf{7}, 9735 (2017) 
- an open access article distributed under the terms of 
the Creative Commons CC BY license (https://creativecommons.org/licenses/), 
which permits unrestricted use, distribution, and reproduction in any medium, provided the original work is properly cited.
}
\end{figure}

To summarize, PP-NEGF enjoys all the standard tools of quantum field theory.
This includes well-developed rules for the diagrammatic technique together with established ways of building conserving approximations.
Like reduced density matrix techniques, PP-NEGF allows access to state-specific information while presenting a simple and controlled approach to account for time-nonlocal (non-Markovian) effects.
Unfortunately, formulation in the extended Hilbert space and the necessity to impose the normalization condition Eq.~\eqref{PPnorm} makes the method more difficult to reliably apply to full counting statistics (FCS).
Only noise (second cumulant) estimates within perturbative expressions have so far been reported the literature.\cite{meir_shot_2002,wu_noise_2010}
Practical applications using PP-NEGF are often limited to the lowest (second) order in the diagrammatic expansion---the non-crossing approximation---which may fail qualitatively and break symmetries even in rather simple limits (see Fig.~\ref{fig4}).
We now turn to discuss another many-body flavor of the NEGF with several advantages.

\subsection{Hubbard NEGF}
Historically, Hubbard Green's functions were introduced for the description of electronic correlations in narrow bands. \cite{hubbard_electron_1967}
Their main usage was in the formulation of perturbation theory around the atomic limit. 
Hubbard NEGFs are a nonequilibrium analog of the formulation, where the isolated molecule is the starting point and (similarly to PP-NEGF) a diagrammatic perturbative expansion is considered in the molecule--contacts coupling.
The main idea is to consider correlation functions between Hubbard operators:
\begin{equation}
\label{Hubop}
 \hat X_{S_1S_2} = \lvert S_1\rangle\langle S_2\rvert.
\end{equation}
These take the form
\begin{equation}
\label{GFHubNEGF}
 G_{(S_1S_2)(S_3S_4)}(\tau_1,\tau_2) = -i \langle T_c\,\hat X_{S_1S_2}(\tau_1)\,\hat X_{S_3S_4}^\dagger(\tau_2)\rangle.
\end{equation}
Similarly to PP-NEGF, Hubbard GFs are constructed around the nonequilibrium atomic limit and yields molecular state-specific information.
The spectral decomposition (\ref{spectral}) allows for a standard GF to be calculated from the Hubbard GF, but the inverse process is not possible.
Contrary to PP-NEGFs, Hubbard GFs live within a physical Hilbert space, allowing for example simulation of FCS.
While standard NEGF contains only information about correlations between transitions and PP-NEGF contains that of states, Hubbard NEGFs include both.

Some of the first applications of Hubbard NEGFs to transport were carried out within the equation-of-motion (EOM) approach, where auxiliary field techniques allow for closed-form equations in terms of the GFs and their functional derivatives to be written. \cite{sandalov_theory_2003,fransson_non-equilibrium_2010}
The method was useful in modeling elastic \cite{fransson_nonequilibrium_2005,sandalov_nonlinear_2006,sandalov_shell_2007} and inelastic \cite{galperin_inelastic_2008} transport,
and was also utilized in ab initio simulations. \cite{yeganeh_transport_2009}
Despite these successes, the somewhat arbitrary nature of the auxiliary fields makes it difficult to generalize these methods into systematic expansions.

Diagrammatic technique for Hubbard GFs was developed in studies of strongly correlated equilibrium lattice models. \cite{izyumov_statistical_1988,ovchinnikov_hubbard_2004}
The technique is based on an observation that (anti-)commutator of two Hubbard operators yields not a number, but another Hubbard operator; and on the equilibrium form of the density operator.
Given these, it can be shown that any Hubbard operator within an $n$-time correlation function can be permuted with other Hubbard operators and with the density operator under the trace, leading to an expression containing only $n-1$-time correlators.
Iterating this procedure generates an expression analogous to Wick's theorem and allows for diagrammatic expansions to be formulated: by resumming diagrams, one then arrives at a Dyson-like equation for so-called locators.
Finally, locators be multiplied by time-non-local factors accounting for the spectral weight and renormalization 
of quasiparticles to obtain the Hubbard GFs. 

In Ref.~\onlinecite{chen_nonequilibrium_2017} this idea was extended to Keldysh GFs.
The generalization assumes (as in NEGF) that in some far-off past time the molecule and contacts were separated and that after transients die out, the system does not remember its initial state.
Thermal equilibrium for the separated molecule and contacts are therefore assumed for the far-off initial state.
With this, each term in the Taylor expansion of the $S$-matrix can be evaluated by using the standard Wick's theorem for bath operators, and the generalized Wick's theorem for Hubbard operators within the molecule.
Later, in Ref.~\onlinecite{bergmann_electron_2019}, it was pointed out that the generalized Wick's theorem can equivalently be formulated for any local density operator that is a function of the molecular Hamiltonian only.
For example, such a density operator results from the Markov Redfield/Lindblad QME, and therefore the Hubbard NEGF technique can be formulated as an expansion starting from an unbroadened QME solution.

\begin{figure}[b]
\centering\includegraphics[width=0.8\linewidth]{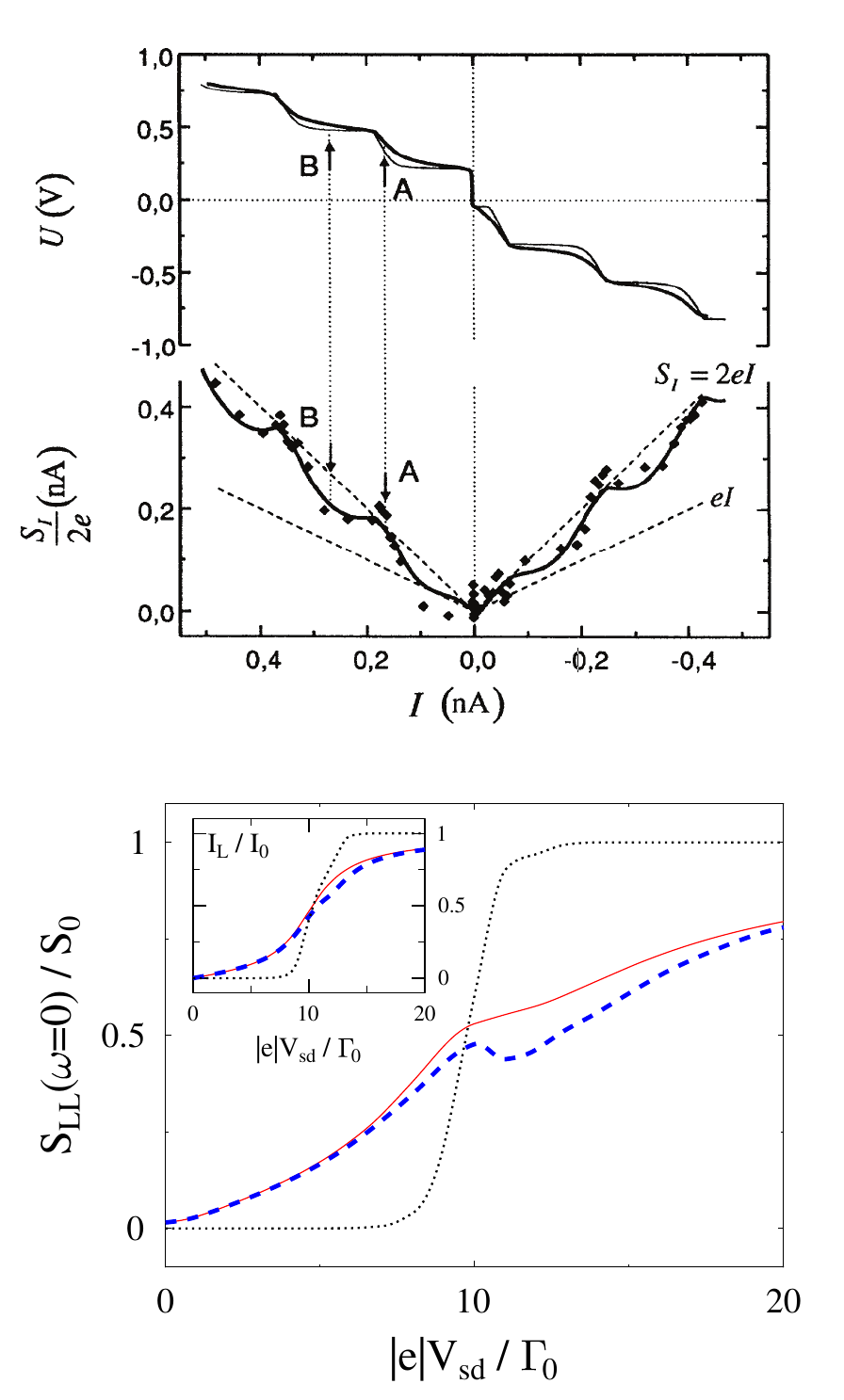}
\caption{\label{fig5}
Shot noise suppression in the Coulomb blockade regime. Shown are experimental data (top) 
and results from NEGF (solid line, red); Lindblad/Redfield QME (dotted line, black); and Hubbard NEGF (dashed line, blue) simulations (bottom).
The inset in the bottom panel shows current--voltage characteristics.
Top figure is reprinted with permission from 
H.~Birk, M.~J.~M.~de~Jong, and C.~Sch{\" o}nenberger, Phys. Rev. Lett. \textbf{75}, 1610-1613 (1995).
Copyright (1995) by the American Physical Society.
Bottom figure is reprinted from K.~Miwa, F.~Chen, and M.~Galperin, Sci. Rep. \textbf{7}, 9735 (2017) 
- an open access article distributed under the terms of 
the Creative Commons CC BY license (https://creativecommons.org/licenses/), 
which permits unrestricted use, distribution, and reproduction in any medium, provided the original work is properly cited.
}
\end{figure}

Hubbard GF diagrammatics can be applied to FCS. \cite{miwa_towards_2017} 
It was shown that it is an inexpensive method capable of reproducing satisfactory noise characteristics in molecular junctions over a wide range of parameters.
In this sense, Hubbard NEGFs perform comparably to standard NEGF at the weakly interacting limit and to QME for strong many-body interactions. 
In the Coulomb blockade regime the second order Hubbard NEGF theory is able to reproduce experimentally observed shot noise suppression, whereas NEGF and QME treatments at the same perturbation order fail (see Fig.~\ref{fig5}).

Another recent application of Hubbard NEGFs is in the derivation of current-induced nuclear forces: in particular, the nontrivial electronic friction force was evaluated. \cite{chen_current-induced_2019,chen_electronic_2019}
Ref.~\onlinecite{chen_current-induced_2019} derives a general expression for the friction, applicable to any form and/or strength of electron--nuclei coupling.
The Hubbard NEGF result was benchmarked with respect to exact simulation in a simple model, and performed well.
The expression obtained appears to be defined by retarded projection of the Hubbard GFs, and contains previous known results as its limiting cases.
For example, the celebrated Head-Gordon and Tully (HGT) expression for electronic friction is shown to be given by a (diagonal) subset of the contributions given by Hubbard GF theory, taken at the noninteracting quasi-particle limit.
The Hubbard GF accurately reproduces the HGT result in the limit of weak molecule--contact coupling.
Based on this developmemt, Ref.~\onlinecite{chen_electronic_2019} discussed the possibility of engineering molecular friction in single molecule nanocavity junctions.

The Hubbard NEGF theory has proven particularly useful in simulations of molecular optoelectronic devices. \cite{galperin_photonics_2017} 
Here, an insulating layer is placed between the molecule and contacts in order to prevent nonradiative energy transfer between them.
The corresponding model parameters are therefore close to the atomic limit.
Formulation in the many-body state description is helpful for avoiding spurious  orbital-shifting and renormalization effects upon charging and/or excitation.
This picture also lends itself to a simple explanation of the difference between transport and optical gaps in junctions, where the two gaps correspond to transitions between different pairs of molecular many-body states.

\begin{figure}[t]
\centering\includegraphics[width=\linewidth]{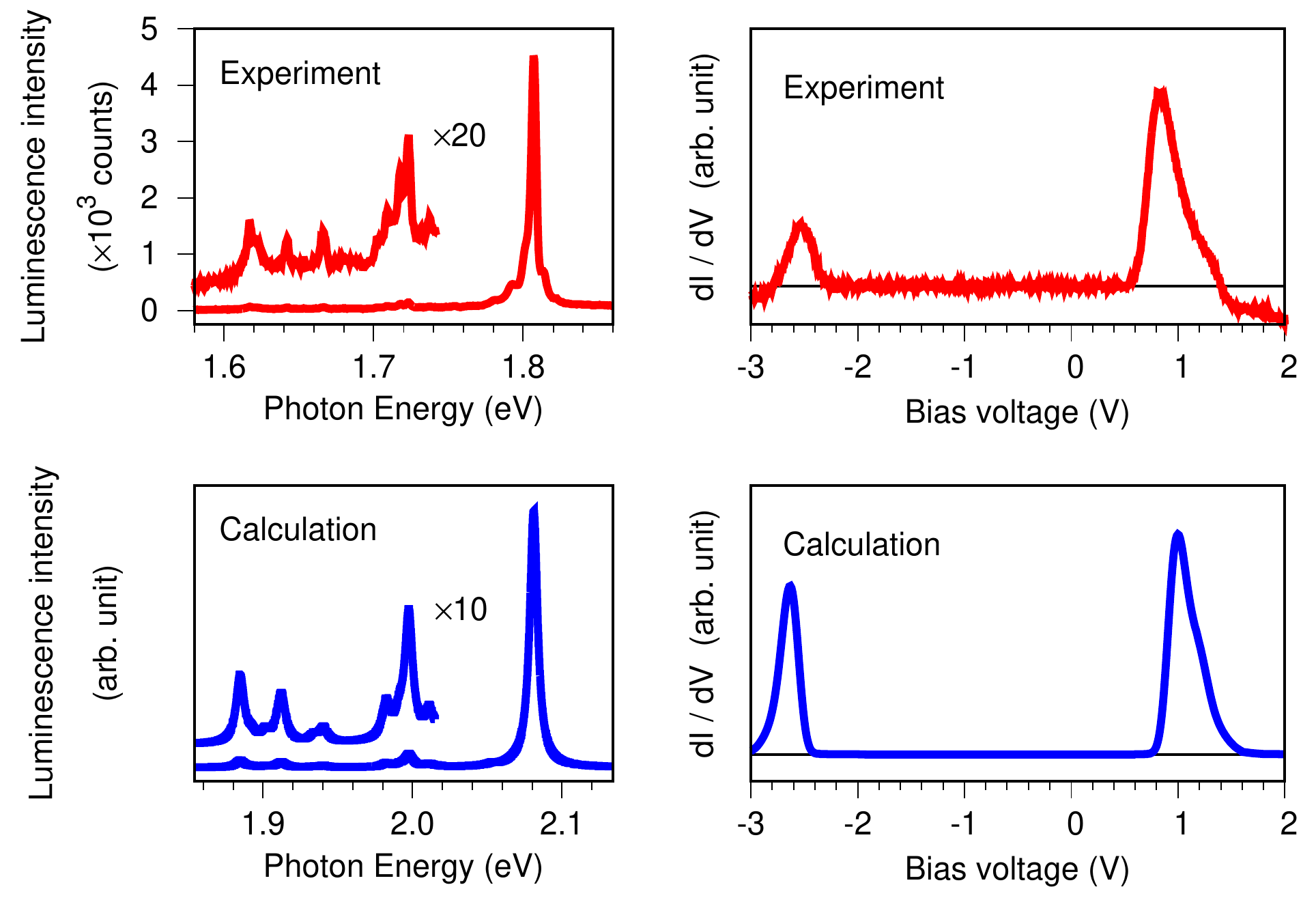}
\caption{\label{fig6}
Optical response in a single-molecule STM junction. Experimental data (red line, top) is shown along with ab initio Hubbard NEGF simulations (blue line, bottom) of the single molecule luminescence (left) and conductance (right). 
Reprinted (adapted) with permission from K.~Miwa, H.~Imada, M.~Imai-Imada et al, Nano Lett. \textbf{19}, 2803-2811 (2019). 
Copyright (2019) American Chemical Society.
}
\end{figure}

Recently, the Hubbard NEGF was employed in ab initio simulations of transport and optical response in molecular junctions.
Ref.~\onlinecite{miwa_hubbard_2019} modeled an experiment dealing with photocurrents induced in nitroazobenzene molecular junctions, and Refs.~\onlinecite{kimura_selective_2019} and \onlinecite{miwa_many-body_2019} respectively, simulated bias-induced phosphorescence and luminescence (see Fig.~\ref{fig6}). 
 
In summary, much like PP-NEGF, the Hubbard NEGF is formulated in the many-body state basis of the isolated molecule.
This makes it useful for treating systems with strong many-body interactions within the molecule and weakly coupled to the baths, such as molecular optoelectronic devices.
The formulation can exactly account for all intra-molecular many-body interactions, and provides a way of incorporating high-level quantum chemistry methods (CI, CCSD, etc.) into the realm of open nonequilibrium molecular junctions.
Recently developed diagrammatic technique for Hubbard NEGFs allows one to take into account system--bath interactions in a controlled manner.
Simultaneously, access to correlation functions of excitations makes Hubbard NEGF similar to the standard NEGF, which may account for the surprisingly high level of accuracy observed at strong system--bath coupling.
Also, contrary to the PP-NEGF, the Hubbard NGEF is constructed within the physical Hilbert space, making it usable for evaluating FCS.
The main difficulty with the Hubbard NEGF is the absence of a clear way to construct an analog of the Luttinger--Ward functional.
Because of this, so far no clear rules for constructing conserving approximations within the Hubbard NEGF have been found.


\subsection{Dual fermion approach}
We have discussed NEGF theory, which is formulated around the noninteracting or quasiparticle limit, and therefore applicable to weak many-body interactions, $U\ll\Gamma$.
We have also discussed PP and Hubbard NEGF; these are formulated around the atomic limit, and suitable for weak system--baths coupling, $U\gg\Gamma$.
The parameter regime where the two characteristic energy scales are similar, $U\sim\Gamma$, has to be treated within an approach which does not rely on existence of small parameter (see Fig.~\ref{gf_regimes}).
Rigorous treatment of such regimes therefore requires numerically exact approaches.
However, such approaches are typically computationally expensive and limited to very simple, minimal models like the Anderson impurity model.
Thus, especially for ab initio purposes, inexpensive approximate methods that remain reasonably accurate in the absence of a small parameter are in high demand.

A new method that is potentially applicable to realistic simulations in the absence of a small parameter is the dual fermion (DF) approach.
Historically, the DF method was formulated for equilibrium lattice models as a way to account for nonlocal corrections to dynamical mean field theory. \cite{rubtsov_dual_2008,antipov_opendf_2015,rohringer_diagrammatic_2018,zhou_nonequilibrium_2019}
However, it was quickly realized that it can serve as a generic ``superperturbative'' method for solving quantum impurity problems, both in \cite{hafermann_superperturbation_2009} and out  \cite{jung_dual-fermion_2012,kehrein_dual-fermion_2012} of equilibrium.

The formulation of the DF method for solving an impurity problem or molecular junction comprises two main parts.
In the first step, a (solvable) reference system containing the many-body interactions is chosen.
The physical system's action, $S[\bar d,d]$, is written in terms of the known action of the reference system $\tilde S[\bar d,d]$ and the difference between the exact hybridization self energy $\Sigma^B$, and that of the reference system $\tilde \Sigma^B$:
\begin{equation}
\begin{split}
S[\bar d,d] &\equiv \bar d_1\,[G_0^{-1}-\Sigma^B]_{12} \, d_2 + S^{int}[\bar d,d]
\\
 & = \tilde S[\bar d,d] + \bar d_1\,[\tilde\Sigma^B-\Sigma^B]_{12}\, d_2.
 \end{split}
\end{equation}
Here, the subscript indices $1,2$ indicate both quantum numbers and contour times, and a sum over repeated subscripts is implied;
$G_0$ is the noninteracting GF of the molecular region $M$; and $S^{int}[\bar d,d]$ is the action resulting from the many-body interactions.

The reference action is interacting, and therefore Wick's theorem does not hold and the standard tools of many-body perturbation theory cannot be directly applied to it.
To resolve this, in the second step, a Hubbard--Stratonovich transformation is employed.
This introduces a new auxiliary quasiparticle, the ``dual fermion'', and makes it possible to formulate a perturbation theory in the coupling between the dual and real fermions.
When the real fermions are traced out, the free dual fermion theory is an exact effective theory reproducing the interacting reference action at its zeroth order.
Corrections to this take the form of effective many-body interactions between the dual fermions, which can now be treated perturbatively.

\begin{figure}
\centering\includegraphics[width=0.9\linewidth]{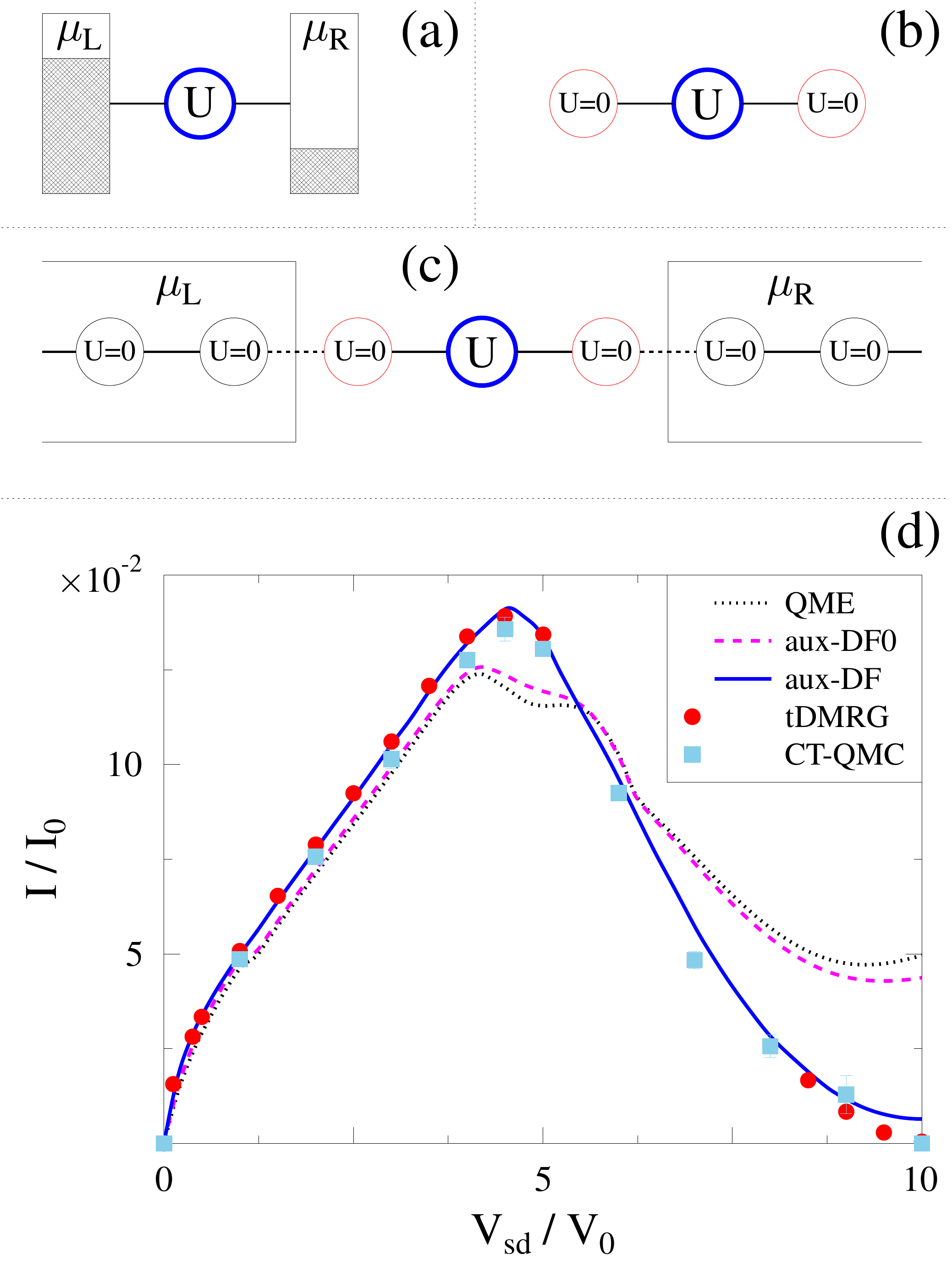}
\caption{\label{fig7}
Current--voltage characteristic. 
Shown are (a) the Anderson impurity model, (b) the reference system within the original DF approach,\cite{jung_dual-fermion_2012}
(c) the reference system within the aux-DF approach \cite{chen_auxiliary_2019} and (d) results of simulations: aux-DF (solid line, blue) is benchmarked against numerically exact tDMRG (circles, red) and Inchworm CT-QMC (squares, blue) results.
Figure is reprinted with permission from
F.~Chen, G.~Cohen, and M.~Galperin, Phys. Rev. Lett. \textbf{122}, 186803 (2019).
Copyright (2019) by the American Physical Society.
}
\end{figure}

In Ref.~\onlinecite{jung_dual-fermion_2012}, which considered the nonequilibrium Anderson impurity model
(see Fig.~\ref{fig7}a), the reference system was chosen to be the interacting impurity site along with one to three bath sites.
This necessarily entails an approximate treatment of the impurity--bath hybridization.
The finite, closed reference system used in Ref.~\onlinecite{jung_dual-fermion_2012} (see Fig.~\ref{fig7}b) does not account for the dissipative dynamics of the open quantum system, or for the nonequilibrium population dynamics, even at the classical level.
It therefore makes it difficult to capture the steady state.
Ref.~\onlinecite{chen_auxiliary_2019} suggested that an auxiliary open system with Markovian leads, described 
by a Lindblad/Redfield QME,  be used as the reference system instead (see Fig.~\ref{fig7}c).
This aux-DF method allows for the problem to be formulated directly in the nonequilibrium steady state and offers substantially improved accuracy in comparison to the closed system DF method.
It also significantly reduces the numerical cost of simulations by avoiding the long time propagation that is required in the original formulation.

The idea of constructing accurate QMEs by explicitly taking into account the dynamics of some bath sites is known in the literature as the auxiliary master equation approach (AMEA), and has been employed as an impurity solver in nonequilibrium dynamical mean field theory. \cite{arrigoni_nonequilibrium_2013,dorda_auxiliary_2014,dorda_auxiliary_2015,dorda_optimized_2017}
This method becomes exact at the limit of an infinite number of auxiliary sites,\cite{tamascelli_nonperturbative_2018,chen_markovian_2019} 
but remains approximate in practice, where only finite number of auxiliary sites can be treated.
The accuracy of AMEA depends on a fitting procedure for the effective hybridization function $\Sigma$
of the original problem in the reference system.
A large number of auxiliary modes is often needed to obtain a high level of precision.
There is therefore a trade-off between accuracy and efficiency: more auxiliary modes enable a better fit, but make the solution of the auxiliary QME more expensive.
In comparison, the alternative and complementary strategy embodied by the aux-DF method is to perform a perturbative expansion in the difference between the hybridization functions of the physical and reference systems.
This can be done for a smaller number of modes, while relying on relatively poor fitting, because the superperturbation theory provides further improvement over the bare AMEA result.
While more expensive than AMEA for a given number of auxiliary sites, aux-DF therefore also yields substantially more accurate results at that same limit (compare solid and dotted lines in Fig.~\ref{fig7}d).

Test simulations of the Anderson impurity model performed in Ref.~\onlinecite{chen_auxiliary_2019} were benchmarked against numerically exact tDMRG and Inchworm continuous time quantum Monte Carlo (CT-QMC) results and demonstrated high accuracy (see Fig.~\ref{fig7}d).
The Inchworm method will be discussed below. 
Note that aux-DF requires only a fraction of the computational cost of the numerically exact techniques to achieve results that (for the parameters tested) are of comparable accuracy.
Like the DF method in equilibrium, aux-DF correctly describes the limits of weak and strong molecule--bath coupling, and allows for interpolation between them.
The superperturbation theory converges rapidly at both limits because it becomes exact when \emph{either} the hybridization or the many-body interactions are small.
Indeed, for weak interactions or strong molecule--bath coupling, the vertices can be neglected and DF reproduces standard NEGF perturbation theory.
For strong interactions (or the atomic limit, where molecule--bath coupling is weak) the DF method is also accurate, due to exactly accounting for the local many-body interactions in the reference system.
Currently, the main obstacle on the road to an ab initio implementation of the aux-DF methodology is in the need to find a numerically inexpensive way to evaluate vertices in the solution of the reference system.

We also note that very recently, a generalization of the method able to address not just electron, but also energy transport was presented and applied to the Anderson--Holstein model: the auxiliary QME--dual boson technique (aux-DB).\cite{chen_nonequilibrium_2019}
This allows to account for electron--electron, electron--phonon or electron--photon interactions while treating particle and energy fluxes on an equal footing.


\subsection{Time-dependent processes with Green's functions}
Molecular electronics is usually focused on steady state properties, which tend to be more experimentally accessible.
Nevertheless, measuring and modeling time-dependent and transient phenomena is an important challenge when considering time-dependent quantum transport, time-resolved photoabsorption, pump-probe type experiments, Auger decay processes and other ultrafast phenomena. \cite{freericks_theoretical_2009,ochoa_pump-probe_2015,covito_real-time_2018,covito_benchmarking_2018,perfetto_cheers:_2018,bostrom_charge_2018}
As such problems typically require GFs explicitly depending on two times, they tend to be more computationally challenging in GF methods: in time-local methods like wavefunction and density matrix approaches, the state is always described by a single time index.
For noninteracting systems, some of the most efficient methods for obtaining GFs actually rely on wavefunction dynamics.\cite{gaury_numerical_2014,gaury_dynamical_2014}
The upside is that GFs also provide rich time-nonlocal information.

One exception to the above rule is problems with periodic driving, which can be relatively easily treated in GFs using Floquet theory \cite{kohler_controlling_2003,kohler_charge_2004,kohler_driven_2005,stefanucci_time-dependent_2008,wu_floquet-greens_2008,tsuji_correlated_2008,wu_kondo_2010,rai_electrically_2013,eissing_functional_2016} 
and similar considerations. \cite{wang_current_1999,park_charge_2011,Sena_Junior_2017}
If two-time GFs do need to be considered, a straightforward approach is to employ discretization on a two-dimensional time grid.
Accounting for the fast decay of GFs with separation of their time variables then allows some simplifications.
Such considerations have been applied to the equations of motion of GFs in both their integral (Dyson) or differential (Kadanoff--Baym) form. \cite{zhu_time-dependent_2005,myohanen_kadanoff-baym_2009,stan_time_2009,my_oh_anen_kadanoff-baym_2010}
Other works perform the time discretization directly on the Keldysh contour. \cite{freericks_nonequilibrium_2006,freericks_nonequilibrium_2006,freericks_quenching_2008,souto_transient_2018,avriller_buildup_2019}
To perform long-time simulations at better computational scaling, approximations can be used.
For example, the generalized Kadanoff--Baym ansatz (GKBA) and other semianalytical ideas have been used. \cite{maciejko_time-dependent_2006,tuovinen_time-dependent_2013,albrecht_long_2013,latini_charge_2014,tuovinen_time-dependent_2014,karlsson_generalized_2018,hopjan_molecular_2018,hopjan_initial_2019}
These effectively reduce the computation to single-time evolution.
For noninteracting NEGF in particular, this can be done in several highly efficient ways without the need for approximations. \cite{gaury_numerical_2014,ridley_current_2015,popescu_efficient_2016,rahman_non-equilibrium_2018,ridley_electron_2019}
A particularly interesting idea that should be noted in this context was proposed in Ref.~\onlinecite{balzer_auxiliary_2014}, where the problem of solving the Dyson equation is mapped onto a noninteracting auxiliary problem with additional degrees of freedom.


\subsection{Quantum Monte Carlo methods}\label{exact}
The past decade has seen a great deal of progress in numerically exact GF methods for simple models of single molecule junctions, in particular with regard to continuous time quantum Monte Carlo methods (CT-QMC).
All such approaches presently remain far too expensive to be usable in an ab initio context, where a large interacting basis must be taken into account and interactions exist throughout both the molecular and lead regions.
Nevertheless, in simple models, they allow for reliable access to genuinely strongly correlated physics, where there is no small parameter, and where interpolation between analytically solvable regimes is insufficient.

The CT-QMC approaches we discuss below have equilibrium predecessors formulated in imaginary (rather than real or Keldysh) time.\cite{gull_continuous-time_2011}
Real time Quantum Monte Carlo approaches to even the most minimal models used to describe nonequilibrium electronic transport through junctions were commonly thought to be limited to short propagation times by the \emph{dynamical sign problem}: an exponential decrease in the signal-to-noise ratio as a function of time.
For the spin--boson model, this view has been challenged by path integral Monte Carlo methods formulated in terms of density matrices and influence functionals.\cite{mak_monte_1996,egger_path-integral_2000}
Two sets of rather different CT-QMC methods have addressed fermionic transport, and will be briefly reviewed below.
One relies on an expansion in many-body interactions and the other on one in molecule--lead hybridization, but both include ideas that are independent of the choice of expansion.

\subsubsection{Interaction expansion}
The real time interaction expansion for nonequilibrium impurity models was introduced in Ref.~\onlinecite{werner_diagrammatic_2009}, and is based on an earlier equilibrium method.\cite{rubtsov_continuous-time_2004}
Confusingly, it is sometimes referred to as the ``weak-coupling'' approach.
The term ``weak-coupling'' here refers to a physical regime of the impurity problem at weak interaction strength, not to the strength of the molecule--lead coupling or hybridization.
However, the interaction expansion is (a) numerically exact, and therefore not limited to a particular regime; and (b) is in fact more efficient in the presence of weak interactions and strong molecule--lead coupling.
Despite featuring a dynamical sign problem, the method was successfully used to explore the transient and steady state current--voltage characteristics of a junction with electron--electron interactions\cite{werner_diagrammatic_2009,werner_weak-coupling_2010} and further employed within dynamical mean field theory.\cite{tsuji_correlated_2008}

More recent advances in real time interaction expansion Monte Carlo stem from an interesting realization about starting from the CT-QMC interaction expansion on the Keldysh contour of Ref.~\onlinecite{werner_diagrammatic_2009}, but then summing explicitly over Keldysh branch indices to obtain a moment expansion.\cite{profumo_quantum_2015,bertrand_quantum_2019}
It turns out that this leads to two crucial simplifications: first, the approximate dynamics simulated by the Monte Carlo process become exactly unitary; and second, the dynamical sign problem plaguing any naive interaction expansion essentially vanishes.
The cost of the summation over branch indices is exponential in the order, such that the expansion must be truncated at a finite but high order in the many-body interaction (so far, orders around 10--15 have been used).
An altered scheme allows for removing this explicit exponential cost at the cost of reintroducing a sign problem, and may lead to a set of improved intermediate techniques.\cite{moutenet_cancellation_2019}

The bare interaction expansion (regardless of how efficiently it is evaluated) has a very limited convergence radius.
However, it was shown that this radius can be extended with the aid of analytical continuation techniques.\cite{profumo_quantum_2015}
In later work, it was demonstrated that understanding the analytical structure of the impurity problem can be leveraged to construct conformal transformations that provide highly accurate analytical continuation procedures that continue to work deep within the strongly correlated regime.\cite{bertrand_reconstructing_2019}
Additionally, statistical noise from the Monte Carlo procedure is larger at large interaction strengths, but can be reduced by post-selecting samples that obey known, exact physical constraints at the limit of infinite interactions.\cite{bertrand_reconstructing_2019}

\subsubsection{Hybridization expansion}
The use of the molecule--lead hybridization expansion in quantum transport began a year earlier than that of the interaction expansion: it was introduced in Ref.~\onlinecite{muhlbacher_real-time_2008}, which considered electron--phonon interactions.
This was based not on CT-QMC, but on an earlier algorithm by Hirsch and Fye.\cite{hirsch_monte_1986}
Here, times along the contour were discretized, and Monte Carlo steps comprised changing the occupation at one discrete interval.
This was rapidly followed by a continuous time approach, which eliminates the need for discretization;\cite{schiro_real-time_2009,albrecht_bistability_2012} and by applications to electron--electron interactions.\cite{schmidt_transient_2008,werner_diagrammatic_2009,schiro_real-time_2010}
An interesting recent development shows that the CT-QMC problem for a model with two spin channels can be mapped onto one with a single channel, resulting in a significant reduction in the sign problem.\cite{kubiczek_exact_2019}

At any finite time, the bare hybridization expansion converges at all interaction strengths and no analytical continuation is required.
Nevertheless, the method is strongly limited by the dynamical sign problem to times that are on the order of the inverse molecule--bath coupling.

\begin{figure}
	\centering\includegraphics[width=\columnwidth]{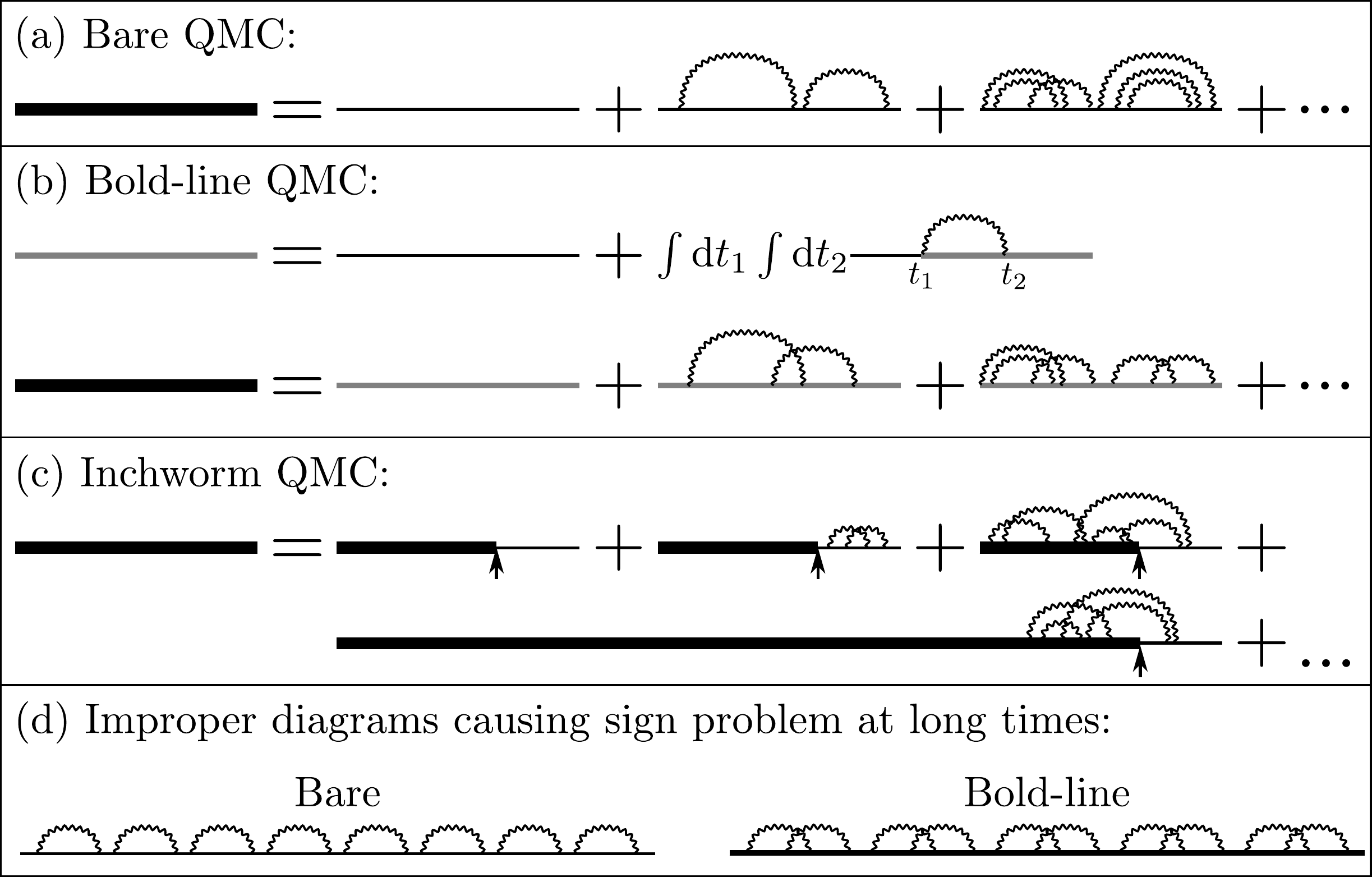}
	\caption{\label{fig8}
		(a) Bare, (b) bold and (c) Inchworm hybridization expansions, along with (d) simple diagrams that contribute to a sign problem in the first two expansions.\cite{cohen_taming_2015}
Figure is reprinted with permission from 
G.~Cohen, E.~Gull, D.~R.~Reichman, and A.~J.~Millis, Phys. Rev. Lett. \textbf{115}, 266802 (2015).
Copyright (2015) by the American Physical Society.
}
\end{figure}

The hybridization expansion Monte Carlo methods are formulated in terms of reduced propagators, much like the propagator techniques mentioned in Sec.~\ref{ssec_pp_negf}.
As illustrated in Fig.~\ref{fig8}a, the expansion for the propagator (thick horizontal line, with contour time going from left to right) can be expressed in diagrammatic notation as a sum over products of isolated molecular propagators (thin horizontal line) and hybridization lines given by molecule--lead GFs (curved wiggly lines).
GFs of any order, as well as density matrix elements in either the molecule or leads, can be expressed in terms of modified propagators with additional operators inserted at the appropriate Keldysh times.

It is possible to perform partial resummations over a contour ``self-energy''; for example, at lowest order this provides the noncrossing approximation for the propagator, as illustrated in the top half of Fig.~\ref{fig8}b and denoted by a ``bold'' or partially thickened gray line.
Bold-line propagators of low orders can be obtained semi-analytically, and then used to construct a dressed expansion containing only corrections to the underlying theory.
These corrections can be summed in a Monte Carlo procedure expressed entirely in terms of the renormalized bold-line propagators (lower half of Fig.~\ref{fig8}b), as first shown in equilibrium.\cite{gull_bold-line_2010}
When applied to real time population dynamics, this idea was shown to greatly reduce the dynamical sign problem, allowing access to timescales several times longer than those accessible by the corresponding bare expansion.\cite{gull_numerically_2011,cohen_numerically_2013}
The bold-line method was then extended to nonequilibrium GFs and used to explore the spectral and current--voltage characteristics of junctions with electron--electron interactions.\cite{cohen_greens_2014,cohen_greens_2014-1,antipov_voltage_2016}

\subsubsection{Inchworm expansion}
The next advance in the field, dubbed ``Inchworm Monte Carlo'', is not specific to the hybridization expansion; however, this was its first application.\cite{cohen_taming_2015}
The main idea is first to take advantage of the fact that propagators over short contour time intervals are easily evaluated; and second to note that propagators obey a kind of contour causality, in the sense that propagators over long time intervals can be efficiently expressed in terms of propagators over shorter time intervals (see Fig.~\ref{fig8}c).
A formally exact series for the exact propagator over some interval $(t,t^\prime)$ can be written in terms of all propagators over intervals $(s,s^\prime)$ with $t\le s, s^\prime \le t_{\uparrow} \le t^\prime$.
At an intermediate stage of the algorithm, the Inchworm time $t_{\uparrow}$, denoted by the arrow in Fig.~\ref{fig8}c, is the time up to which all propagators are known.
By performing the summation over all diagrams contributing to the expansion, propagators up to the longer time $t^\prime$ are obtained.
A series of such ``inching'' steps eventually generates the propagator over the entire contour and physical observables, but---crucially---propagators over long intervals are only calculated when propagators over slightly shorter intervals are already known and taken advantage of.
Therefore, each inching step has a relatively small sign problem, which in practice does not grow exponentially with time.\cite{cohen_taming_2015}
As an example (see Fig.~\ref{fig8}d), a set of trivial diagrams that are generated by low order self energies can easily be shown to generate an exponential dynamical sign problem in both the bare (left) and bold-line (right) expansions.\cite{cohen_taming_2015}
Diagrams of this form are summed implicitly within the Inchworm expansion, such that this particular source of sign problems is completely removed.

The Inchworm hybridization expansion can be truncated in diagram order for improved efficiency, though this is not necessary in all regimes: an Inchworm expansion of order $n$ converges to a dressed expansion for the propagator self energy up to that order, and often converges very rapidly.
No analytical continuation is necessary.
When higher order contributions (number of vertex pairs $n \gtrsim 10$, corresponding to order ~20 self-consistent perturbation theory) are needed, fast summation techniques become useful.\cite{boag_inclusion-exclusion_2018}
Nevertheless, the method is based on a resummed perturbation theory.
If the order needed for convergence at a given set of parameters is too high, it eventually fails.
Furthermore, the computational cost of existing implementations is effectively quadratic in the propagation time, whereas the interaction expansion of Ref.~\onlinecite{profumo_quantum_2015} can provide results at very long times with no additional expense.

Whereas Ref.~\onlinecite{cohen_taming_2015} only considered population dynamics in interacting quantum junctions, the method was quickly extended to currents and GFs,\cite{antipov_currents_2017} and used to explore the transient and steady state properties of voltage-driven interacting junctions in the strongly correlated regime.\cite{krivenko_dynamics_2019}
The Inchworm method also proved to be suitable to the numerically exact evaluation of FCS.\cite{ridley_numerically_2018}
This was used to explore the effect of junction geometry on electronic current and its fluctuations,\cite{ridley_lead_2019} and provides access to thermal transport properties and entropy generation.\cite{ridley_numerically_2019}

We briefly note that the Inchworm idea is not limited to the real time hybridization expansion in quantum transport: it has also been applied within dynamical mean field theory;\cite{dong_quantum_2017} shown to be effective for combating the dynamics sign problem in the spin--boson model using two different diagrammatic expansions;\cite{chen_inchworm_2017,chen_inchworm_2017-1,cai_inchworm_2018} and very recently, to be applicable to the imaginary time sign problem in multiorbital impurity models.\cite{eidelstein_multiorbital_2019}


\section{Conclusions}\label{conclude}

We presented a short pedagogical review of methods utilized in the study of quantum transport through molecular junctions.
Three broad categories were considered: wavefunction based methods, density matrix and Green's function (GF) formulations.
While the approaches are equivalent when treated exactly, in practice exact treatment is only possible for noninteracting systems where the single particle picture is accurate.
For interacting problems, each category of methods leads to different approximations and numerically exact schemes, with widely varying advantages and disadvantages.
In wavefunction methods, observables and dynamics can be expressed by quantities taken at a single time, but the amount of information scales exponentially with the size (e.g. number of orbitals) of the complete system.
This includes both the molecule and contacts.
In density matrix methods, at the cost of introducing either non-Markovian dynamics of single-time observables or a Markovian approximation, the spatial region contributing to the exponential scaling is reduced to that of the interacting molecular region.
In GF methods, at the cost of having to address time-nonlocal dynamics of two- or multi-time correlation functions, the exponential spatial scaling can be completely removed.
Methods from all categories have seen rapid progress in recent years, and modern approaches can efficiently mix and match between categories to best suit specialized problems.
Theoretical considerations were illustrated with numerical examples in model and ab initio simulations, and promising avenues for future research were pointed out.

The main focus of this review is on GF approaches.
GFs often provide an efficient and accurate way to simulate experimentally measurable observables.
This includes electronic and energy currents carried by either fermionic charge carriers or bosonic modes.
It can also include current fluctuations and full counting statistics.
On the theoretical side, GF methods provide a methodological framework for constructing controlled approximation schemes around solvable limits.
We discussed and compared standard NEGF, which expands around the noninteracting limit; its many-body flavors (PP- and Hubbard NEGF), formulated around the atomic limit; and superperturbation theory (dual-fermion and dual-boson) that can provide accurate predictions in both limits, and bridges between them.

We also discussed numerically exact GF schemes based on Quantum Monte Carlo algorithms, that can provide reliable results at all parameter regimes, though only for very simple models and at a large computational cost.
Until recently it was thought that such methods are limited to very short propagation times in nonequilibrium systems, due to dynamical sign problems.
However, new methodological frameworks introduced in the last few years, such as Inchworm Monte Carlo, have shown that this sign problem can often be circumvented.
Such diagrammatic Monte Carlo methods are based on perturbative expansions similar or identical to the ones used in approximate GF techniques.
However, the Monte Carlo algorithms allow for systematically summing all corrections in an unbiased manner up to very high, or sometimes effectively infinite, orders.

In the context of molecular electronics, we noted that diagrammatic perturbation theory around the noninteracting NEGF is convenient
for treating molecular junctions with relatively weak many-body interactions, such as intra-molecular electron--electron interactions.
This is often the case when considering, e.g., transport through $\pi$-conjugated organic polymers.
On the other hand, the PP- and Hubbard NEGF diagrammatic expansions are perturbative in the molecule--bath coupling strength rather than the interaction.
This makes them useful, e.g, in the presence of strong electron--electron interactions within the molecule, as is typical in molecular optoelectronic devices.
Dual techniques, a newer alternative, are relatively inexpensive universal impurity solvers that can remain accurate even in regimes where neither simple perturbative expansion works well.
Each of these methods has, or could potentially have, a numerically exact technique associated with it that can tackle nonperturbative regimes.

The theory of nonequilibrium transport through single molecule junctions is at a turning point.
The community has developed an impressive array of numerical and analytical techniques that can address a wide variety of physical problems, and progress is still being made.
We hope this review will be helpful to readers in navigating through the forest of available methods and using them to push our scientific understanding and technological mastery of these systems to new heights.
Data sharing is not applicable to this article as no new data were created or analyzed in this study.

\begin{acknowledgments}
MG acknowledges support by the National Science Foundation (CHE-1565939).
GC acknowledges support by the Israel Science Foundation (Grant No. 1604/16).
\end{acknowledgments}


%

\end{document}